# SOLARNET Metadata Recommendations for Solar Observations

*Version 2.2 – 18 March 2024*

*Stein Vidar Hagfors Haugan and Terje Fredvik*

This document has received funding from the European Union's Horizon 2020 and FP7 programmes under grant agreements No 824135 and 31295. Version 2.0 of this document was the final version produced under these grants, but it keeps evolving as further needs and issues arise. The latest *published* version of this document can be found at https://arxiv.org/abs/2011.12139, but the *latest working copy* of the document can be found at https://sdc.uio.no/open/solarnet/. Please use that version when adding comments/suggested changes (using track changes), before sending to prits-group@astro.uio.no

## *Abstract*

Until the advent of the SOLARNET recommendations, metadata descriptions of Solar observations have been standardized for space-based observations, but the standards have been mostly within a single space mission at a time, at times with significant differences between different mission standards. In the context of ground-based Solar observations, data has typically not been made freely available to the general research community, resulting in an even greater lack of standards for metadata descriptions. This situation makes it difficult to construct multi-instrument archives/virtual observatories with anything more than the most basic metadata available for searching, as well as making it difficult to write generic software for instrument-agnostic data analysis. This document describes the metadata recommendations developed under the SOLARNET EU project, which aims foster more collaboration and data sharing between both ground-based and space-based Solar observatories. The recommendations will be followed by data pipelines developed under the SOLARNET project and others see footnote for a full list (may need updating)[1]. These recommendations are meant to function as a common reference to which even existing diverse data sets may be related, for ingestion into solar virtual observatories and for analysis by generic software.

[1] Full list per July 2021: Solar Orbiter SPICE, SST CHROMIS/CRISP (SSTRED), AISAS/Lomnicky Stit COMP-S/SCD, SAMNET, Gregor HiFI/GFPI, ROB USET, Alma pipeline for Solar data (SOAP), INAF IBIS-A (IBIS data Archive), PADRE/MeDDEA

Table of Contents (if sub-appendices appear wrong, delete the "…." to the left of appendix title)

# *Part A. Description of FITS keywords*

# 1. About file formats

The most common practice in the solar remote sensing community is currently to use the FITS Standard file format for disseminating solar remote sensing observations. For this reason, this document describes how to include the metadata content through keywords inside FITS files, but *that does not preclude the use of other file formats*. In many ways, this document simply uses FITS notation as a language to express the underlying metadata requirements.

For a discussion about file names and how to group observational data between or inside different files, see Appendix V: Other recommendations.

# 2. Header and Data Units (HDUs) in FITS files

FITS files may contain one or more Header and Data Units (HDUs) of different types, e.g., primary HDUs, image extensions, and binary table extensions, containing data and a header with metadata stored as keyword-value pairs. Primary HDUs and image extensions are for almost all practical purposes identical: The primary HDU should simply be regarded as "*the first HDU, which must exist and is always an image HDU but is not required to contain any data, just a header*". Thus, all SOLARNET recommendations applying to an extension HDU also applies to the primary HDU.

This document primarily describes how the SOLARNET recommendations apply when using primary and image HDUs to store observational data, but Appendix IV also explains how the recommendations may be applied to observational data stored in binary table extensions. In this context, observational data are data values derived from solar photons recorded by a detector. Other types of data, e.g., temperatures, voltages, atmospheric conditions etc., will be regarded as auxiliary data.

In this document, HDUs storing observational data will be referred to as Obs-HDUs.

There are many keywords that are not mentioned in this document which have an established definition in the FITS Standard, Papers I-V, Thompson (2006), other references, and existing or past projects. As far as possible, such keywords should be used when appropriate, and should not be used in conflict with that definition. New keywords should not be invented if there is already a keyword in established use that covers the needs of the keyword. In general, see the "Other sources of keywords with established use" from the References section before inventing a new keyword.

By default, FITS string values are limited to 68 characters, but the CONTINUE Long String Convention (see References) may be used in order to allow keywords to contain strings longer than 68 characters. *Note that this convention must not be used with any of the mandatory or reserved keywords defined in the FITS standard*.

In order to prevent having to specify keyword information that is common to all HDUs in a file, keyword inheritance may be used according to the FITS Header Inheritance Convention (see References).

This document introduces three new mechanisms that are not part of the FITS standards, but may be useful in fully describing observations: Appendix I explains how to describe keywords that vary as a function of WCS coordinates, Appendix I-d explains how to pin-point and (optionally) associate values to specific pixels/locations inside a data cube, and Appendix III explains how to signal that an HDU is a part of a larger set of HDUs (e.g., a time series) contained in multiple files.

## 2.1. Naming of HDUs in SOLARNET FITS files

All HDUs – including the primary HDU – in SOLARNET FITS files *must* contain the string-valued keyword **EXTNAME**, and each **EXTNAME** value must be unique within the file[2]. **EXTNAME** must not contain the characters comma or semicolon except as prescribed for the variable-keyword mechanism (Appendix I), the pixel list mechanism (Appendix I-d) and the meta-observation mechanism (Appendix III). In addition, **EXTNAME** must not start with a space, but any trailing spaces are ignored. Finally, the CONTINUE Long String Keyword Convention must not be used with **EXTNAME**, since this is a reserved keyword defined in the FITS standard.

## 2.2. Fully and partially SOLARNET-compliant Obs-HDUs

All fully SOLARNET-compliant *and* partially SOLARNET-compliant Obs-HDUs *must* contain (in addition to all mandatory FITS standard keywords) the following mandatory keywords (this also applies to the primary HDU if it is an Obs-HDU):

**EXTNAME**
**SOLARNET**
**OBS_HDU**
**DATE-BEG**

Obs-HDUs cannot contain keywords with definitions in conflict with other SOLARNET-defined keywords unless they occur in a comma-separated list in the keyword **SOLNETEX**. This mechanism may sometimes be necessary to ensure backwards compatibility with existing utilities. Keywords listed in **SOLNETEX** will be ignored by SOLARNET-aware utilities. The **SOLNETEX** mechanism must *not* be applied to FITS standard keywords.

The **SOLARNET** keyword is used to signal if an Obs-HDU is fully SOLARNET-compliant (**SOLARNET=1**) or partially SOLARNET-compliant (**SOLARNET=0.5**).

Fully SOLARNET-compliant Obs-HDUs *must* contain all mandatory SOLARNET/FITS standard keywords described in Section 15 that apply (depending on the nature of the observation) and *must not* have any of those mandatory keywords listed in **SOLNETEX**.

Partially SOLARNET-compliant Obs-HDUs need not contain all keywords summarised in Section 15.

Both fully and partially SOLARNET-compliant Obs_HDUs *must* have **OBS_HDU=1**.

---

[2] There is one exception to the SOLARNET rule requiring unique EXTNAMEs: As per the official FITS WCS mechanism for **Lookup** coordinate distortions, all extensions containing lookup tables must have **EXTNAME='WCSDVARR'**. These extensions must instead be distinguished by having different values of the seldom used FITS Standard **EXTVER** keyword.

## 2.3. Other HDUs

Other HDUs in the same file as an Obs-HDU may be used to store additional data that is required to describe the observations, to allow instrument-specific utilities to function correctly, to interpret the data correctly, or to enable further calibrations to be made. Specific cases of such HDUs are variable-keyword HDUs (Appendix I), pixel list HDUs (Appendix I-d) and Meta-HDUs (Appendix III).

Obs-HDUs that are neither fully nor partially SOLARNET-compliant may use the mechanisms described in 0, Appendix I-d or Appendix III, if they have **SOLARNET=-1**. In fact, the HDUs described in these appendices may themselves use these mechanisms, if they have **SOLARNET=-1**.

## 2.4. Comments

We strongly urge pipeline designers to use the FITS mechanism for commenting keywords (using a forward slash after the keyword value). Although pipeline designers may know all their cryptic keywords by heart while writing the pipeline, this will not be the case a year or more later.

Any units for keyword values should be enclosed in square brackets at the beginning of the keyword comment (Section 4.3 of the FITS Standard).

We are considering making a machine- and human-readable catalogue of keyword comments used in this document, making it possible to create utilities that fill them out automatically for SOLARNET-defined keywords. However, the main utility of such a wrapper would be to also identify keywords with no comment at all, or comments without unit specifications (where appropriate).

Additional comments may be added using the **COMMENT** keyword or by leaving the keyword field blank – see Section 4.4.2.4 in the FITS Standard.

# 3. The World Coordinate System (WCS) and related keywords

The World Coordinate System (WCS) is a very comprehensive standard that should be used for the description of physical data coordinates in Obs-HDUs.

In some earlier data sets, the data coordinates are not specified using the WCS standard, but rather through e.g., **XCEN**, **YCEN**, **FOVX**, and **FOVY**, etc. Future pipelines, however, should *only* use the full, recommended WCS standard, without any deprecated features (e.g., **CROTAi**) or any instrument- or mission-specific practices[3].

All keywords described in this Section are defined by the FITS Standard and Papers I-V. See also Thompson (2006).

[3] If a full description seems impossible through the existing WCS framework, please contact prits-group@astro.uio.no.

## 3.1. Fundamental WCS coordinate specification

As a reference, the most commonly used conversion from the set of pixel indices *($p_1$, $p_2$ $p_3$, … $p_N$)* to a physical WCS coordinate $c_i$ is given by the following formula:

$$c_i(p_1, p_2, p_3, \dots, p_N) = \mathrm{CRVAL}i \; + \; \mathrm{CDELT}i \sum_{j=1}^{N} \mathrm{PC}i_j\left(p_j - \mathrm{CRPIX}j\right)$$

Thus, to calculate the physical (world) coordinates $c_i$ at any point within the data cube, the following entities are involved:

- **CRPIXj** values specify the data cube pixel indices for a "reference point".
- **CRVALi** is the value of world coordinate **i** at this reference point.
- **PCi_j** is a linear transformation matrix between data cube dimension **j** and coordinate axis **i**, which can be used to specify rotations, shear, transpositions and reflections.
- **CDELTi** is the increment of world coordinate **i** at the reference point.

Here, **j** goes from **1** to the number of data dimensions, i.e., from **1** to **NAXIS**, and **i** goes from **1** to the number of physical coordinates. The number of physical coordinates is normally equal to **NAXIS** but may be larger or smaller as optionally specified in **WCSAXES**. Cases in which **WCSAXES** are used to indicate a different number of dimensions than **NAXIS** could be e.g.:

- when trailing singular dimensions are being suppressed in the writing of the file, as happens in IDL.
- when there are more coordinates that vary throughout the data cube than there are data cube dimensions, e.g., in a raster scan with (x,y,lambda) coordinates, with a time coordinate that is also a function of the x coordinate.
- when dummy coordinates are used in table lookup of coordinates, in order to minimise the storage space requirements.

**CTYPEi** is used to specify the nature of the coordinates and their projections. In solar observations, the most appropriate values include **HPLN-TAN** and **HPLT-TAN** (solar coordinates; Thompson 2006), **WAVE** (wavelengths in vacuum; Paper III), **UTC** (time; Paper IV), and **STOKES** (Stokes parameter; Paper I).

Coordinates may also be given in table lookup form (Section 6 in Paper III), for use with e.g., Fabry-Pérot imaging spectroscopy with uneven spacing in the wavelength direction. Also, WCS even allows for specifications of distortions down to a pixel-by-pixel level basis if required (see Paper V).

In ground-based observations, image restoration techniques such as MOMFBD leave behind apparent local movements of image features. Such residual effects represent local errors/distortions in the coordinate system specified by the HDU's WCS keywords. In the FITS Standard Section 8.2, it is specified that a "representative average" of such random errors may be given in the keywords **CRDERi** (for axis number **i**).

Likewise, representative averages for systematic errors in the coordinates may be given in the keywords **CSYERi** (for coordinate number **i**). Thus, **CSYERi** should be used to represent the uncertainty in the pointing/position of the image as a whole, and uncertainties in the wavelength calibration for spectrometric data.

If a coordinate system has been determined or refined through the use of some external reference image(s) or other source(s), or even been adjusted manually, the keyword **PRREFn** should be used to give a comma-separated list of the images/sources/people, see Section 8. If it is not possible to give specific image names/references, the name of the instrument, filter, etc. should be given. Since such images must obviously be (near) co-temporal with the data in the Obs-HDU, this should not introduce much ambiguity.

WCS allows *multiple* coordinate systems to be specified for each data cube. Alternate coordinate systems are defined by specifying additional sets of WCS keywords, each set with a letter **A-Z** appended at the end. Keywords that may be used in this way are sometimes indicated with a suffix **a**, e.g., **CTYPEia** (see the FITS Standard Section 8.2.1). As an example, **CTYPE1B** specifies the type of coordinate **1** in coordinate system **B**.

In particular, multiple coordinate systems can be used to correctly describe data such as rasters, with one system describing the spatial-wavelength coordinate system (*x, y, lambda*), and another describing the temporal-spatial-wavelength coordinate system (*time, y, lambda*). Imaging observations scanning through the wavelength dimension could have a primary system describing (*x, y, lambda*) and a second coordinate system (*x, y, time*).

However, in many such cases it is simpler and more appropriate to use a single coordinate system with four coordinates, e.g., (*x, y, lambda, time*). This comes naturally if the observations are repeated and concatenated in time (i.e., resulting in a 4-dimensional data cube), but can also be used when scans are stored individually, (i.e., as 3-dimensional data cubes). In such cases, it is necessary to specify the number of coordinates with the keyword **WCSAXES=4** in order to account for the time coordinate that is not represented by a dimension in the data cube.

For rotating FOVs in a time series, the table lookup algorithm for coordinates (see Paper III Section 6) must be used, with a joint table lookup of coordinates **HPLN-TAN**, **HPLT-TAN** and **UTC**. Since table lookup of WCS coordinates is performed with linear interpolation, it is normally possible to represent such a rotating FOV with a coordinate table that has size **(x,y,t)=(2,2,t)**, where **t** may be significantly smaller than the number of time steps in the time series. For highly non-linear rotation rates the indexed form of the table lookup algorithm may be used to vary the sampling of the FOV coordinates with time.

For observations (instruments) where the plate scale/pointing is derived from measurements of the apparent solar radius versus the physical size, the keywords **RSUN_REF** should be used to report the reference value for the physical radius used in the calculations (see Thompson 2010, Section 8).

For descriptions of distortions of coordinates in complex data sets, e.g., cavity errors, see Appendix VI.

## 3.2. WCS positional keywords and relative radial velocity

Ground based observatories must report their geographical location using the keywords **OBSGEO-X**, **OBSGEO-Y**, and **OBSGEO-Z**, implicitly stating that the observer is following Earth rotation (see Precision effects for solar image coordinates within the FITS world coordinate systems, Section 3; Paper III Section 7). In principle, the coordinates should be given in ITRF geocentric coordinates. However, for SOLARNET purposes, GPS coordinates are an acceptable proxy.

Earth-orbiting satellites must report their position through **GEOX_OBS**, **GEOY_OBS**, and **GEOZ_OBS**. Contrary to the **OBSGEO-X/Y/Z** keywords, these keywords do *not* implicitly imply that the coordinates are fixed w.r.t. Earth's rotation, but are otherwise identically defined (ITRF, but GPS is an acceptable proxy). For many observations, these keywords must be reported using the variable-keyword mechanism (Appendix I) since the spacecraft might move considerably during the observation.

For deep space missions, the keywords **DSUN_OBS** (distance from Sun centre in metres), **HGLN_OBS** (longitude), and **HGLT_OBS** (latitude) must be used to report the instrument position in the Stonyhurst Heliographic system (see Thompson 2006, Sections 2.1 and 9.1). The distance from the Sun centre in astronomical units may be reported in **DSUN_AU** (in addition to **DSUN_OBS**). Note that the Solar B angle is identical to **HGLT_OBS**, and although it is a duplication of information, it may be reported also in **SOLAR_B0** for convenience.

If other coordinate systems or positional information are given for the observer position, they should follow the specifications in Thompson (2006), Sections 2.1 and 9.1.

For spectrometers (and for some narrow-band imagers), the radial velocity between the instrument and the Sun may be important. Unfortunately, WCS does not have a mechanism for specifying this without also correcting the wavelength scale to account for the Doppler shift (see Paper III, Section 7). Such a correction is not traditionally applied in FITS files within the solar physics community. To specify that no such wavelength correction has been done, **SPECSYS** must be set to **'TOPOCENT'** and **VELOSYS** must be set to 0.0. In order to specify the observer's radial velocity relative to the Sun, the non-WCS keyword **OBS_VR** (given in m/s) must be used (possibly as a variable keyword). Positive velocities are outward from the Sun (i.e., **OBS_VR**=*dr/dt*).

As mentioned also in Section 2, many keywords already established elsewhere but not mentioned in this document may apply. Such keywords should never be used in conflict with established use. In particular, see "Other sources of keywords with established use" under References. A few that are related to those defined in this section are: **SOLAR_P0** (apparent angle from observer location between celestial north and solar north), **SOLAR_EP** (apparent angle from observer location between celestial north and ecliptic north), **RSUN_ARC** (apparent photospheric solar radius in arc seconds), and **CAR_ROT** (Carrington rotation number for the reference pixel pointed to by **CRPIXj** values).

# 4. Time-related WCS keywords

**DATE-BEG** *must* be given, referring to the start time of the data acquisition in the time system specified by the **TIMESYS** keyword, which has the default of **'UTC'**. The **TIMESYS** value applies to all **DATE-** keywords, **DATEREF** (see Section 4.1), and several other date-valued keywords.

**DATE-END** may be given, referring to the end of data acquisition.

DATE-AVG may be used to give the average date of the observation. However, there is no unambiguous definition of the average when applied to observations with varying cadence or varying exposure times.

Note that we do *not* recommend using the DATE-OBS keyword mentioned in the FITS Standard, since this is not explicitly defined there, and has a history of somewhat ambiguous use (see Paper IV).

The observer's position may be important when comparing the times of observations from different vantage points – in particular when at least one of the observations is space based. Thus, the keywords DSUN_OBS, HGLN_OBS, and HGLT_OBS (Section 3.2) may be important w.r.t the timing of the observations.

## 4.1. Specifying WCS time coordinates

The literature describing all the possible methods of specifying WCS time coordinates is very complex, but except in unusual circumstances, the following prescription should be sufficient:

CTYPEi='UTC' should be used as the name of the WCS time coordinate. However, applications should also recognize the value 'TIME' as having the same meaning, for historical reasons.

Also, DATEREF *must* be set to the zero point of the WCS time coordinate. I.e., for pixels that have the CTYPEi='UTC' coordinate equal to zero, the time is the value given in DATEREF. In most cases the values of DATEREF and DATE-BEG will be identical, but note that *according to the FITS standard,* DATE-BEG *is not a default value* for DATEREF, thus DATEREF may not be omitted. The existence of both keywords allows e.g., midnight to be used as a zero point for the time coordinate for multiple observations recorded during the following day, each having different values of DATE-BEG.

# *5. Description of data contents*

A description of the actual data contents is important for the interpretation of an observation. Such a description is also important for finding relevant observations in an SVO.

## 5.1. Data type/units (BTYPE/BNAME/BUNIT)

The keywords BTYPE, BNAME, and BUNIT should be used to describe the nature of the data. The notation of mathematical expressions in BUNIT and BNAME should follow the rules in Table 6 of the FITS Standard, e.g. "log(x)" is defined as the common logarithm of x (to base 10).

BUNIT should be used to indicate the units of the values in the data cube.

BTYPE should be used to describe what the data cube itself represents. This keyword is not mentioned in any FITS standard document, but it is a natural analogy to the CTYPEi keywords used to indicate the WCS coordinate type. When possible, we recommend using the Unified Content Descriptors (UCD) version 1+ (see References) when specifying BTYPE, or gice the UCD description in a separate UCD keyword. It may be that the UCD scheme does not cover all data types encountered in solar observation. Thus, it may be necessary for the solar community to decide upon other values for this keyword. This is currently an unresolved issue.

**BNAME** may be used to provide a human readable explanation of the data contents. This keyword is not mentioned in any FITS standard document, but it is a natural analogy to the **CNAMEi** keywords used to provide additional description of the WCS coordinate.

## 5.2. Exposure time, binning factors

The exposure time used in the acquisition of an Obs-HDU should be given in the keyword **XPOSURE** - not in **EXPTIME**. The reason why **EXPTIME** should not be used is that in *some cases* it has been used for individual exposure times in summed multi-exposure observations, introducing an ambiguity. According to the recommendation in Paper IV, **XPOSURE** should always contain the *accumulated* exposure time whether or not the data stems from single exposures or summed multiple exposures.

When the data are a result of multiple summed exposures with identical exposure times, the keywords **NSUMEXP** and **TEXPOSUR** can be used to indicate the number of summed exposures and the single-exposure time, respectively.

When the **XPOSURE** or **TEXPOSUR** values vary as a function of time or any other of the Obs-HDU's dimension(s), the variable-keyword mechanism can be used to specify their exact values as a function of those dimensions (see Appendix I for further details). This would typically be the case when Automatic Exposure Control is used - both **XPOSURE** and **TEXPOSUR** could vary as a function of time.

Note that if the data has been binned, the **XPOSURE** keyword should reflect the *physical* exposure time, not the sum of exposure times of the binned pixels. Binning should be specified by the keywords **NBINj**, where **j** is the dimension number (analogous to the **NAXISj** keywords). E.g., for an observational data cube with dimensions (**x,y,lambda,t**) where 2x2 binning has been performed in the **y** and **lambda** directions (as is sometimes done with slit spectrometers), **NBIN2** and **NBIN3** should be set to 2. The default value for **NBINj** is 1, so **NBIN1** and **NBIN4** may be left unspecified.

In order to provide a simple way to determine the combined binning factor (for archive searches), the keyword **NBIN** should be set to the product of all specified **NBINj** keywords.

## 5.3. Cadence

Cadence may be a very important search term. A meta-Obs-HDU may be used to report such attributes even if it is impossible to do so in the constituent HDUs (Appendix III).

The planned/commanded cadence (frame-to-frame spacing measured in seconds) should be reported in **CADENCE**. The average (actual) cadence should be reported in **CADAVG**.

The cadence *regularity* is also important: The keywords **CADMAX** and **CADMIN** should be set to the maximum and minimum frame-to-frame spacing. **CADVAR** should be set to the variance of the frame-to-frame spacings.

Some instruments take interleaved observation series with a difference in cadence between different filters ("A" and "B"), e.g., AAABAAAB. For such a series, **CADENCE** for the A series should be the planned *median* spacing between A exposures.

For e.g., on-going synoptic observation series stored with single exposures in separate files (thus separate HDUs) it may be impossible to use the Meta-observation mechanism. The **CADENCE** keyword should be set to the planned series' cadence. The rest of the keywords should be set based on the available history of the synoptic series.

## 5.4. Instrument/data characteristics etc.

In order to characterise the spectral range covered by an Obs-HDU, the keywords **WAVEMIN** and **WAVEMAX** should be used to specify the minimum and maximum wavelengths.

The magnitude of the wavelength related keywords mentioned in this section (**WAVExxx**) must be specified in **WAVEUNIT**, given as the power of 10 by which the metre is multiplied, e.g., **WAVEUNIT=-9** for nanometre. We recommend that **WAVEUNIT** corresponds to the **CUNITi** value of the WCS wavelength coordinate, if any, e.g., if **CUNITi='Angstrom'** then **WAVEUNIT=-10**.

**WAVEREF** should be set to **'air'** or **'vacuum'** to signal whether wavelengths are given for air or vacuum. We recommend that **WAVEREF** corresponds to the **CTYPEi** value of the WCS wavelength coordinate, if any. E.g., if **CTYPEi='AWAV'** then **WAVEREF ='air'**.

For spectrometers, the **WAVEMIN/WAVEMAX** values represent the range of wavelengths covered by the Obs-HDU. If the file contains multiple readout windows, the wavelength coverage of the entire file may be reported in **WAVECOV='(<WAVEMIN1>-<WAVEMAX1>, <WAVEMIN2>-WAVEMAX3>, …)'**

For filter images, the definition is somewhat up to the discretion of the pipeline constructor since effective response curves are never a perfect top-hat function. Bear in mind that these two keywords are primarily meant to be used for search purposes. E.g., if someone wants an observation covering a specific wavelength lambda, the search can be formulated as "**WAVEMIN < lambda < WAVEMAX**". In other words, it might be wise to include more than the "intended" or "nominal" min/max wavelengths of a filter: sometimes parts of an extended tail should be included if it covers a potentially interesting emission line that is normally very weak but may be strong under certain conditions. We suggest that the wavelengths at which the response function is 0.1 times the peak might be a good choice, unless other considerations make other choices more appropriate. This should be based on a measured response function if available – otherwise it should be based on a design specification or theoretical basis. We reiterate, though, that the criteria are up to the discretion of the pipeline designers. The criteria used to set these keywords should in all cases be specified in the keywords' comment.

For filter images, the **WAVELNTH** keyword may be set to the "characteristic wavelength" at which the observation was taken. For EUV imagers, this keyword typically identifies the most prominent emission line in the bandpass. For a spectrometer **WAVELNTH** might also be the middle of the wavelength range of the HDU, but we leave the exact definition up to the pipeline designers.

In addition, the keyword **WAVEBAND** may be used for a human-readable description of the waveband, typically the (expected) strongest emission/absorption line in HDUs containing spectrometer observations (or specifying the continuum region), or the most dominant contributing line in filter images.

For radio observations, **BNDCTR** may be used instead of **WAVELNTH** to specify a corresponding frequency in Hz.

For filter observations where a more thorough specification of the response curve is required for a proper analysis or for search purposes, the variable keyword **RESPONSE** may be used – see Appendix I.

The **RESPONSE** keyword should also be used for spectrometers where there are significant variations in the response across the dataset.

If the data has already been corrected for a variable response, the response function that has been applied should instead be given in the variable keyword **RESPAPPL**.

For spectrometric data, the resolving power R should be given in the keyword **RESOLVPW**. For slit spectrometers, the slit width in arc seconds should be given in **SLIT_WID**.

### *5.4.1. Polarimetric data reference system*

Different Stokes values are normally stored together in a *single* extension, with a **STOKES** dimension and an associated **STOKES** coordinate to distinguish between the different values (**I/Q/U/V** or **RR/LL/RL/**etc). The **STOKES** coordinate should vary along the **STOKES** dimension according to Table 29 in the FITS Standard. Pixels containing I, Q, U, and V values should have **STOKES** coordinates 1, 2, 3, and 4, respectively. If different Stokes values are stored in different extensions or in different files, the **STOKES** coordinate should still be specified - either as a "phantom" WCS coordinate without an associated data dimension (i.e., **WCSAXES > NAXIS**) or as a regular coordinate for a singular data dimension.

Existing conventions for specifying the *reference system* for Stokes vectors use celestial coordinates (**RA**/**DEC**), but for Solar observations this is not practical. Thus, we define here that SOLARNET-compliant FITS files should use a right-handed reference system *(x, y, z)* with the *z* coordinate oriented either parallel or antiparallel with the line of sight towards the observer. The axes must be explicitly specified by the keyword **POLCCONV** in the form **'(+/-x, +/-y, +/-z)'** where **x**, **y**, and **z** are valid WCS coordinate names. E.g., **POLCCONV='(+HPLT,-HPLN,+HPRZ)'** means that the reference system's *x* axis is parallel to the **HPLT** axis (Solar North), and *y* is *antiparallel* to the **HPLN** axis, with *z* pointing towards the observer.

If the polarimetric reference frame is not aligned with any set of WCS coordinate names, a rotation of the reference frame given in **POLCCONV** can be specified in **POLCANGL**. The rotation, specified in degrees, should be applied to the **POLCCONV**-specified system around its third axis. The rotation is counter-clockwise as seen from a point with a positive third-axis coordinate value, taking the sign from **POLCCONV** into account. I.e., specifying a positive angle with **POLCCONV='(…, …, +HPRZ)'** specifies a counter-clockwise rotation as seen from Earth, whereas with **POLCCONV='(…, …, -HPRZ) '** would specify a clockwise rotation as seen from Earth.

## 5.5. Quality aspects

Many quality aspects of ground-based observations change rapidly, even from one exposure to the next. Keywords that describe such quality aspects must therefore often use the variable-keyword mechanism to specify the time evolution of such values, see Appendix I. This mechanism may be used to specify quality-related values for single exposures, average or effective values for composite images, while also allowing an average or effective scalar value to be given in the header.

Until now, there has been little effort in order to characterise quality aspects of ground-based observations in a manner that is *consistent* between different telescopes, and even between different setups at the same telescope. In FITS files from ESO (European Southern Observatory), the keyword **PSF_FWHM** is used to give the full width at half maximum in arc seconds for the point spread function. However, this quantity is generally not available for solar observations. Some adaptive optics systems, however, may record parameters like the atmospheric coherence length $r_0$. If available, the value of $r_0$ should be stored in the keyword **ATMOS_R0**. Since there are multiple ways of measuring this value, its only use should be to reflect the quality of the observing conditions whenever the measurements are performed in the same (or similar enough) way.

If you have suggestions for consistent methods of measuring parameters describing the spatial resolution of observations (or a proxy for it), please contact prits-group@astro.uio.no, so that we can include this method in a later version of the document.

The keyword **AO_LOCK** should be used to indicate the status of any adaptive optics. When specified for individual exposures, the value should be either 0 or 1, but as mentioned above, averages may also be specified as appropriate.

The keyword **AO_NMODE** should be used to indicate the number of adaptive optics modes corrected. As mentioned above, averages may also be specified as appropriate. The type of the modes (e.g., Zernike, Karhunen-Loeve, etc.) should be given in the keyword comment.

The keyword **FT_LOCK** is used to indicate the status of any feature tracking **FT_LOCK=0** (no feature tracking lock) or **FT_LOCK=1** (feature tracking lock) for individual exposures, with appropriate averages as mentioned above.

The keyword **ROT_COMP** should be set to 1 if automated solar rotation compensation was in effect during the observation, and to 0 if not. The keyword **ROT_MODL** should be set to specify the rotation model used for rotation compensation[4]. It can refer to specific, predefined models such as **ALLEN** (Allen, Astrophys. Quantities, 1979), **HOWARD** (Howard *et al.*), **SIDEREAL**, **SYNODIC**, **CARRINGTON**, **SNODGRASS** or **aaa.a** – arcseconds per hour (units [arcsec/h]). See also the SolarSoft routine **diff_rot.pro**. If other models have been used, please contact prits-group@astro.uio.no, or set **ROT_MODL** to **FORMULA**, and specify the formula in the keyword **ROT_FORM**. The formula specified is meant to be human-readable, not machine readable, thus e.g., '**A sin(….)**', using parameter names that are common within your community. An explanation in the comments may be useful. The units should be degrees per day. More important, though is that the coordinate variation is reflected in the WCS description of the data, using cross terms between the time coordinate and spatial coordinates in the **PCi_j** matrix or by tabulating the coordinates.

If other relevant keywords seem necessary, we recommend using keywords starting with '**ROT_**', but please contact us as well.

**ELEV_ANG:** This keyword should be used to quote the telescope's elevation angle at the time of data acquisition, in degrees.

[4] This might be important when comparing observations where cross-correlation cannot be used for alignment - e.g., coronal observations vs. photospheric observations. In such cases, different rotation models might cause a drift between the two. The information in this keyword can be used to prevent misunderstandings and misinterpretations in in such situations.

In some cases, lossy compression has been applied to the data. Depending on the type of compression, different quality aspects will be introduced that should somehow be specified. Since any significant on-board processing should be considered as a processing step in the pipeline, lossy compression may be listed using the **PRxxxxn** keywords described in Section 8.

However, for searching and sorting purposes it would be useful to have a generic numeric keyword describing the loss of quality due to lossy compression.

**COMPQUAL** could therefore be set to a number between 0.0 and 1.0, where 1.0 indicates lossless compression (if any) and 0.0 indicates "all information is lost". In practice, however, the actual value is not crucial, as long as a higher value corresponds to a higher data quality. If there is a choice between different compression algorithms for this instrument, the name of the algorithm should be given in **COMP_ALG** – starting with either **'Lossy'** or **'Lossless'**, then typically a concatenation of all instrument-specific compression-related keywords, separated with slashes.

**OBS_LOG:** Location of the log file that is relevant to this observation, if available, given as a URL.

**COMMENT:** May be used to include the relevant parts of the **OBS_LOG,** and any other relevant comments about the HDU that may be useful for the interpretation of the data.

## 5.6. Data statistics

It may be useful to have statistics about the data cube of a Obs-HDU in order to search for "particularly interesting" files (or to filter out particularly *uninteresting* files for that matter).

**DATAMIN** – the minimum data value
**DATAMAX** – the maximum data value
**DATAMEAN** – the average data value
**DATAMEDN** – the median data value
**DATAPnn** – the **nn** percentile (where **nn** is e.g., **01, 02, 05, 10, 25, 50, 75**, **90**, **95**, **98**, and **99**).
**DATANPnn** – **DATAPnn/DATAMEAN**, i.e., normalized percentiles
**DATARMS** – the RMS deviation from the mean, sqrt( sum( (x-avg(x))^2 )/N )
**DATANRMS** – **DATARMS**/**DATAMEAN**, i.e., normalized RMS.
**DATAMAD** – the mean absolute deviation, avg( abs( x-avg(x) ) )
**DATANMAD** – **DATAMAD**/**DATAMEAN**, i.e., normalized MAD
**DATAKURT** – the kurtosis of the data
**DATASKEW** – the skewness of the data

Note that the calculation of these keywords should be based only on pixels containing actual observational data – not including e.g., padding due to rectification, etc.

### *5.6.1. Missing and saturated pixels, spikes/cosmic rays, padding, etc.*

In some data sets, the data in the HDU may be affected by missing/lost telemetry, acquisition system glitches, cosmic rays/noise spikes, or saturation, hot/cold pixels etc. Some keywords are useful to find/exclude files based on how many such pixels there are. In order to allow such searches, the following keywords should be used:

**NTOTPIX** – the number of *potentially* usable pixels: the number of data cube pixels minus **NMASKPIX**
**NLOSTPIX** – the number of lost pixels due to telemetry/acquisition glitches
**NSATPIX** – the number of saturated pixels
**NSPIKPIX** – the number of identified spike pixels

**NMASKPIX** – the number of dust-affected/hot/cold/padded pixels etc.
**NAPRXPIX** – the number of pixels with approximated values (used by e.g., SolO/SPICE)
**NDATAPIX** – the number of usable pixels: **NTOTPIX - NLOSTPIX - NSATPIX - NSPIKPIX**

Corresponding percentages relative to **NTOTPIX** should be given in **PCT_LOST, PCT_SATP, PCT_SPIK, PCT_MASK, PCT_APRX** and **PCT_DATA**.

It is strongly recommended that this naming pattern is followed whenever there is a need to specify further "classes of pixels". I.e., to introduce the pixel class **'SOME'**, **PCT_SOME** should be used to give the corresponding percentage relative to **NTOTPIX**. Analogously, the associated list of pixels (see below), should be named **SOMEPIXLIST**.

### *5.6.2. Explicit listing of missing, saturated, spike/cosmic ray pixels etc.*

Bad pixels may be handled in one of three ways: they can be left untouched, they can be filled with the value of **BLANK** (integer-valued HDUs) or *NaN* (floating-point-valued HDUs), or they can be filled in with estimated values.

For some purposes, it may be useful to keep lists of individual bad pixels or ranges of bad pixels using the pixel list mechanism, see Appendix I-d. This is especially important when the pixels have been filled in with estimated values, storing the original values in the pixel list. Pixel lists that flag individual lost, approximated, saturated, spike or masked pixels, should have **EXTNAME**s equal to **LOSTPIXLIST**, **APRXPIXLIST**, **SATPIXLIST**, **SPIKPIXLIST**, or **MASKPIXLIST** respectively. Original values (when appropriate) should be given in the pixel list's attribute column with **TTYPEn='ORIGINAL'** – see Appendix I-d. for details. For cosmic ray/spike detection, a confidence level (between 0.0 and 1.0) may also be given in an attribute column with **TTYPEn='CONFIDENCE'**. In order to ensure unique **EXTNAME**s for pixel lists belonging to different Obs-HDUs, the pixel list **EXTNAME**s may have a trailing "tag", see Appendix I-d. Pixel lists with other **EXTNAME**s than **LOSTPIXLIST** etc. may of course be used for other purposes, e.g., storing the pixel indices and classification of sunspots, the latter stored as a string valued attribute.

# *6. Metadata about affiliation, origin, acquisition, etc.*

The keywords in this section describe metadata regarding the origin, acquisition, and affiliation of the data. Although not generally required for the *use* of the data, such metadata are very useful w.r.t. e.g., searching, grouping, counting, and reporting. Some of the keywords will not make sense for all data sets, because the nature and nomenclature of the observational scenarios vary. In such cases, leave them out. Also, some of the keywords will have different meanings within different settings, in many cases based on tradition.

We therefore refrain from giving explicit instructions on the usage of many of the keywords. An SVO should allow searching on such keywords by asking for "observations where **PROJECT=xxx**", but it should also be possible to search for "observations where **xxx** occurs in any of the keywords mentioned below".

In general, all keywords below may contain comma-separated lists when necessary. In some cases, it may be a good idea to use both the full name and an acronym.

We *strongly* recommend that all such "free-text" keywords are filled in from lists of predefined texts, strictly controlled by each individual pipeline/instrument team. Experience has shown that free-text fields will be filled in incredibly inconsistently, even the writer's own name. Of course, it would be even better if a community-wide service could be established to homogenise such controlled lists, but this may never happen.

**PROJECT:** Name(s) of the project(s) affiliated with the data
**MISSION:** Typically used only in space-based settings (e.g., the SOHO or STEREO mission)
**OBSRVTRY:** Name of the observatory
**TELESCOP:** Name of the telescope.
**TELCONFG:** Telescope configuration.
**INSTRUME:** Name of the instrument.
**CAMERA:** Name of the camera (**CAM-xxxx** is recommended for other camera keywords).
**GRATING:** Name of the grating used (when there are more than one available).
**FILTER:** Name(s) of the filter(s) used during the observation.
**DETECTOR:** Name of the detector.
**OBS_MODE:** A string (from a limited/discrete list) uniquely identifying the mode of operation.
**OBS_DESC:** A string describing the observation, e.g., “Sit and stare on AR10333”. Should be identical to **OBSTITLE** when no more suitable value is available.
**OBSTITLE:** A more generic/higher-level description, e.g., “Flare sit-and-stare”, “High cadence large raster”), which often corresponds to **OBS_MODE**, though not necessarily as a one-to-one relationship. Used by IRIS and SPICE, corresponds to Hinode **OBS_DEC**. Should be identical to **OBS_DESC** or **OBS_MODE** when no more suitable value is available.
**SETTINGS:** Other settings – numerical values can be given as '**parameter1=n, parameter2=m**'.
**OBSERVER:** Who acquired the data.
**PLANNER:** Observation planner(s).
**REQUESTR:** Who requested this particular observation.
**AUTHOR:** Who designed the observation
**CAMPAIGN:** Coordinated campaign name/number, including instance number, when applicable.

Note also this catch-all keyword:
**DATATAGS:** Used for any additional search terms that do not fit in any of the above keywords.

# *7. Grouping*

*It is very important for an SVO to be able to group search results in a meaningful way!*

E.g., if a search matches 1000 Obs-HDUs, but they are part of only 5 different observation series, it makes sense to have a grouping mechanism to collapse the result listing into only 5 lines, showing some form of summary of the underlying Obs-HDUs for each series.

## 7.1. Pointing ID *separates* observations into groups

To make such grouping work, the concept of a “pointing id” has proven to be useful in e.g., the Hinode archive – it serves to *separate* otherwise identical observations into groups in a logical way.

We therefore introduce the keyword **POINT_ID**, to be given a new, unique string value (e.g., a string giving the date and time of the repointing) every time the telescope is *significantly repointed* - not counting feature tracking or rotation compensation.

Note that changes in the size of the FOV is not considered a repointing. E.g., an alternating series of large-FOV and small-FOV observations should share a single `POINT_ID` value, even if the FOVs are not centred on the same spot – the small-FOV and large-FOV observations may be sorted into separate, parallel groups based other characteristics given by keywords defined below.

Thus, `POINT_ID` is used to *separate* files that should *not* be lumped together in a group because doing so would disguise or misrepresent “what has been observed”.

The exact criteria used for changing the `POINT_ID` value are left up to the pipeline designers/observers, but *we would like to stress the importance of this particular keyword for SVO/archive purposes. Without it, the archive may have no other option than to list your observations with one line per file/Obs-HDU!*

Planning tools/databases are often good sources for of `POINT_ID` values, but with some processing some information may be available from automated logging of telescope orientation, etc.

Even fixed-pointing instruments should use `POINT_ID` to e.g., separate multiple contiguous sequences with breaks in between.

## 7.2. Further separating observations with identical pointing ID

Grouping observations solely on the basis of their `POINT_ID` values will lump together observations regardless of other characteristics such as filters, exposure times, slit widths, etc. In order to separate such disparate observations, further information needs to be taken into account. As an example, grouping on the basis of a concatenation of the values of `POINT_ID`, `CAMERA`, `DETECTOR`, `GRATING`, `FILTER`, `SLIT_WID`, `XPOSURE`, `OBS_MODE`, `SETTINGS`, `NBIN1`, `NBIN2`, `NAXIS1`, and `NAXIS2` would in most cases ensure that only completely identical observations are grouped together (if `XPOSURE` is a measured quantity with slight variations due to e.g., shutter speed variance it should be left out). Exactly which keywords should be used is of course highly instrument dependent.

For other purposes, however, a coarser grouping might make more sense. E.g., omitting `FILTER`, `SLIT_WID`, and `XPOSURE` may still make sense in an archival context if showing a single search result line with a single image gives the user a good (enough) idea of what the observation group contains (e.g., a sunspot or no sunspot). Thus, a hierarchy of groupings can be established, from fine-grained to coarse.

To guide archive designers in the process of presenting grouped observations, we introduce the keywords `SVO_SEPn`, with `n` spanning from `1` to `5`. `SVO_SEP1` should have a comma-separated list of keywords giving the most fine grained grouping, and `SVO_SEP5` should have a comma-separated list of keywords giving the coarsest sensible grouping. Normally, the first keyword should be `POINT_ID`, and it may be the only keyword. Not all `SVO_SEPn` keywords need to be defined, but they should be populated starting with `SVO_SEP1`.

Examples:

```
SVO_SEP1= 'POINT_ID,INSTRUME,DETECTOR,FILTER,NBIN' / Most fine grained separation
SVO_SEP2= ‘POINT_ID,INSTRUME,DETECTOR,FILTER’ / Img. shows target even w/binning
SVO_SEP3= ‘POINT_ID,INSTRUME,DETECTOR’ / Target identifiable in all filters
SVO_SEP4= ‘POINT_ID,INSTRUME’ / Still useful
```

Note that the **SVO_SEPn** keywords are only *guidelines* for archive designers. Typically, an archive will also separate groups by data level if it contains more than a single level.

## 7.3. Group ID ties together logical groups of observations

**SVO_GRP** may be used to achieve the opposite, to tie observations *together* even if they have different **POINT_IDs** and/or have other differing characteristics. Heuristically, a shared value for **SVO_GRP** signals that "if you are interested in this observation, you are probably interested in all of these observations (with the same **SVO_GRP** value)".

## 7.4. Mosaics

For mosaic observations, the keyword **MOSAICID** can be used to tie individual observations together.

# 8. Pipeline processing applied to the data

The concept of "data level" is often used to label data with a particular degree of processing, from raw data up to complex data products.

However, definitions of data levels are extremely instrument-/mission-/pipeline-dependent, and not very useful in terms of explaining what processing has been applied. This concept is useful, however, to ensure that files with different processing level have unique file names. For this reason, the keyword **LEVEL** may be used, as a string value to capture sub-levels such as quick-look versions.

**VERSION** should be set to the *processing version of the file*, an integer that should be increased whenever a reprocessing is performed in order to improve the data set (e.g., with a better flat-field, better detection of cosmic rays, more complete telemetry, etc). The version numbers in files published through an SVO may increase by more than one for each new published "generation", allowing the use of intermediate values for internal/experimental use.

**ORIGIN** should be set to a character string identifying the organization or institution responsible for creating the FITS file. **DATE** should be set to the date of creation of the FITS file.

In addition to the **LEVEL, VERSION** and **ORIGIN** keywords, we recommend that some additional keywords are used in order to indicate the processing steps that has been applied to the data. The four keywords described in Section 8.1 may be used instead of or in addition to the more complex set of keywords described in Section 8.2.

## 8.1. Basic description of processing software

The name and version of the processing software should be specified by those of the following keywords that might apply:

```
CREATOR = 'ZUN_MOMF_PIPELINE'  / Name of software pipeline that produced the FITS file
VERS_SW = '2.5'                / Version of software applied
HASH_SW = 'a7ef89ad998ea7feef4bbc0bbc1bbc2bbc3bbc4' / Commit hash of software applied
VERS_CAL= '2.4'                / Version of calibration pack applied
```

In addition, **PRSTEPn** should specify the nature of the processing steps, if any, that has been applied to the data. Each **PRSTEPn** may contain a comma separated list if multiple processing steps are inseparable. The number **n** specifies the step number and should reflect the order in which the steps have been performed, e.g.:

```
PRSTEP1 = 'FIXED-PATTERN,FLATFIELDING' / First two (inseparable) processing steps
PRSTEP2 = 'CALIBRATION'                / Second processing step
PRSTEP3 = 'DISTORTION-CORRECTION'      / Third processing step
```

Below is a list of recommendations for descriptions of processing steps. If desirable, further specifications may be added, e.g., instead of **'LINE-FITTING'** one may want to use **'GAUSSIAN-LINE-FITTING'** versus **'VOIGT-LINE-FITTING'**. Note that distortion corrections come in two flavours: applied to the data (regridding) or applied to the coordinates. In the latter case, **COORDINATE** should be a part of the processing step description. If you need to add to this list, please contact prits-group@astro.uio.no.

```
ATMOSPHERIC-INVERSION
BIAS-CORRECTION
BINNING
CALIBRATION
CEILING
COMPRESSION
CONCATENATION
DARK-SUBTRACTION
DEMODULATION
DEROTATION
DESPIKING
DESTRETCHING
EDGE-DETECTION
FILTERING
FIXED-PATTERN-REMOVAL
FLATFIELDING
FLOORING
LINE-FITTING
MOMFBD
MULTIPLICATION
PIXEL-FILLING
PIXEL-LEVEL-OFFSET-SUBTRACTION
RADIOMETRIC-CALIBRATION
ROUNDING
SHACK-HARTMANN-DECONVOLUTION
SPATIAL-ALIGNMENT
SPATIAL-COORDINATE-CORRECTION
SPATIAL-COORDINATE-CORRECTION-X
SPATIAL-COORDINATE-CORRECTION-Y
SPATIAL-COORDINATE-DISTORTION-CORRECTION
SPATIAL-DISTORTION-CORRECTION
SPECKLE-DECONVOLUTION
SPECTRAL-ALIGNMENT
SPECTRAL-COORDINATE-CORRECTION
SPECTRAL-COORDINATE-DISTORTION-CORRECTION
SPECTRAL-DISTORTION-CORRECTION
STOKES-INVERSION
SUBTRACTION
SUMMING
TELEMETRY-PARSING
THRESHOLDING
WFS-DECONVOLUTION
```

## 8.2. Detailed description of all processing steps

Each processing step may be described in further detail using some or all of the following keywords in addition to **PRSTEPn**: **PRPROCn**, **PRPVERn**, **PRMODEn**, **PRPARAn**, **PRREFn**, **PRLOGn**, and **PRENVn**.

Libraries used in processing step **n** may be described using some or all of the keywords **PRLIBna**, **PRVERna**, **PRHSHna**, and **PRBRAna**, where **a** is an optional but highly recommended letter **A-Z** to signal that the keyword describes a library, not the main procedure, and to distinguish between different libraries. E.g.:

```
PRSTEP1 = 'MOMFBD  '           / Processing step type
PRPROC1 = 'zun_momf.pro'       / Name of procedure performing PRSTEP1
PRPVER1 =                 1.5 / Version of procedure PRPROC1
PRMODE1 = 'BALANCED'           / Processing mode of PRPROC1
PRPARA1 = 'ITER=5,MANUAL=1'    / List of parameters/options for PRPROC1
PRREF1  = 'miss.influencer@esa.int' / Factors influencing PRSTEP1
PRLOG1  = ' % Program caused arithmetic error: Integer divide by 0' / PRPROC1 log
PRENV1  = '  Kernel: Linux                                                    &'
CONTINUE  '  Kernel release number: 3.10.0-1160.36.2.el7.x86_64               &'
CONTINUE  '  OS: Red Hat Enterprise Linux Server release 7.9 (Maipo)          &'
CONTINUE  '  CPU: Intel(R) Xeon(R) CPU E5-2630L v4 @ 1.80GHz                  &'
CONTINUE  '  IDL 8.5 (Jul  7 2015), memory bits: 64, file offset bits: 64     &'
CONTINUE  ''  / Processing environment of PRSTEP1

PRLIB1A = 'ZUNRED     '        / Software library containing PRPROC1
PRVER1A =               32214 / Version of PRLIB1A
PRHSH1A = 'a7ef89ad998ea7feef4bbc0bbc1bbc2bbc3bbc4' / GIT commit hash for PRLIB1A
PRBRA1A = 'production'         / GIT/SVN repository branch of PRLIB1A

PRLIB1B = 'SSW/vobs/ontology/idl/gen_temp,SSW/packages/sunspice/idl/atest,SSW/&'
CONTINUE  'so/spice/idl/atest,SSW/vobs/gen/idl,SSW/soho/gen/idl/util,SSW/gen/i&'
CONTINUE  'dl_libs/astron/coyote' / Software library containing PRPROC1
PRVER1B =               59549 / Modified Julian date of last mirroring of PRLIB1B
```

In this example, the **zun_momf.pro** routine is part of the **ZUNRED** library and relies on SolarSoft library routines. If further libraries had been used in processing step 1, they would be specified in **PRLIB1B**, etc. Libraries should be listed in the order they appear in the path. Unfortunately, some libraries such as SolarSoft contain internal routine shadowing, in which case each conflicting sub-package must be listed as a separate library in the order they appear in the effective IDL path, i.e., in the system variable **!PATH**[5].

If a single procedure performs multiple steps, it is ok to list each step separately, using the same value in e.g., **PRPROC1** and **PRPROC2**, but different values for **PRSTEP1** and **PRSTEP2**.

The version keywords **PRPVERn** and **PRVERna** should be numerically increasing with increasing maturity of the pipeline. When using libraries with no (numeric) version numbers, the Modified Julian Day (MJD) of the time the library was last mirrored/changed could be used as a version number. Note that for git repositories, it is possible to construct a consistently increasing version number by counting the total number of commits in a fiducial master repository. If none of the above is available, the MJD of the processing itself may be used, implying that the processing uses the latest version available at that date.

---

[5] For SolarSoft, the correct order in which these sub-packages should be listed may be found by tracing the effects on **!PATH** from the beginning (the **$SSW_INSTR** environment variable) to the end, taking into account the effects of **rm_path** and **add_path** statements.

`PRHSHna` is a convenience keyword to easily find the exact git commit for a library that has been used. Note that *it does not replace* `PRVERna`, because `PRHSHna` is not a number that increases with the maturity of the libraries (but such a number can be constructed, see above).

`PRMODEn` is meant for pipelines that may be run with different trade-offs between e.g., signal to noise ratio versus spatial resolution or contrast. This should already be apparent from the other keywords, but `PRMODEn` provides a much simpler way of identifying data processed in a particular way (e.g., "`BALANCED`" or "`HIGH CONTRAST`"). Note that a single observation may be registered multiple times in an SVO with different values of `PRMODEn` - but then a `PRMODEn`-specific identifier in the file name is necessary in order to ensure uniqueness.

The `PRPARAn` keyword can be used in different ways:

1. As a plain comma-separated listing of parameters and their values, signalled by the first character matching the regular expression `'[A-Za-z_$]'`
2. As a JSON string, signalled by the first character being an opening curly bracket
3. As an XML specification, signalled by the first character being a `'<'`
4. As the `EXTNAME` of an ASCII table, signalled by the first character being a `'['`

When `PRPARAn` is used as a reference to an ASCII table, the extension's first three columns should be the parameter name, the parameter value, and an optional comment. It is allowed to use a single ASCII table extension for all processing parameters, but this is of course only possible if all parameter names of all processing steps are distinct. Also, it is recommended to repeat *all* `PRxxxxn` keywords from the Obs-HDU in the header of the ASCII table extension, so that the entire processing history of the data can be seen from that header alone. When parameters are given in separate extensions for each step, it may be less confusing if only the `PRxxxxn` keywords for that particular step is included.

For plain parameter lists, IDL parameter syntax must be used. For ASCII table extensions, values should be specified with regular FITS syntax. For ASCII table extensions, complex values can be written with parentheses containing the real and imaginary parts. Array-valued parameters can be written as a comma-separated list in square brackets in both of these formats.

`PRENVn` can be used to specify the operating environment of the pipeline such as the hardware (CPU type) and the operating system type/version, compiler/interpreter versions, compiler options, etc.[6] The default value of a `PRENVn` keyword is the value of `PRENVn` in the previous processing step, so for a pipeline that has been run from beginning to end in a single environment, only **PRENV1** will have to be specified.

---

[6] This may seem like overkill, but there are instances where e.g., OS versions have mattered, see https://www.i-programmer.info/news/231-methodology/13188-python-script-invalidates-hundreds-of-papers.html, leading to papers being retracted/corrected. Also, for the BIFROST code, on some particular platform a particular CPU instruction optimization has to be turned off with a compiler flag to produce correct results. In the SPICE project it was observed that calculations of mean, variance, skewness, and kurtosis using the built-in IDL function `MOMENT()` differed by as much as 0.6%, 1.6%, 2.4% and 3%, respectively! This may be very significant if such parameters are used to make cuts in a data set.

**PRREFn** is a catch-all keyword that can be used to specify other factors/inputs influencing a processing step, e.g., references to images used for pointing adjustments. **PRREFn** may be a comma separated list of multiple factors/inputs[7].

**PRLOGn** can be used to include a processing log in case messages/warnings from the processing may be of importance.

For some data sets, it may be desirable to include information about how the calibration data has been created/processed (e.g., acquisition of flat fields/dark images). In such cases, the same mechanism should be used, even though the observational data in the HDU is not altered by that processing in itself. The processing steps for the calibration data should have a lower **n** than those steps that use the calibration data (e.g., **PRSTEP1='CALIBRATION-PREPARATION'** and **PRSTEP2='CALIBRATION'**).

# *9. Integrity and administrative information*

The **DATASUM** and **CHECKSUM** keywords (see the Checksum Keyword Convention) should be set in all HDUs to allow a check on whether the data file has been modified from the original or has become corrupted. However, their values in a Meta-HDU (see Appendix III) will be recomputed when constituent HDUs have been combined into a single HDU (after checking the constituent HDUs **DATASUM** and **CHECKSUM**).

**INFO_URL** should point to a human-readable web page describing "everything" about the data set: what it is, how to use it, links to e.g., user guides, instrument/site/telescope descriptions, descriptions of caveats, information about data rights, preferred acknowledgements, whom to contact if you have questions, and repositories of observing/engineering logs.

Upon ingestion of (meta)data into an SVO, the material pointed to by **INFO_URL** and **OBS_LOG** (Section 5.5) might be "harvested" and preserved in such a way that it is possible to retrieve a copy even if the original source is no longer available. It might be possible for an SVO to recursively harvest pages/documents and even auxiliary data such as flat fields being linked to from **INFO_URL**. The harvesting will have to be restricted somehow - presumably limited to links pointing beside or below **INFO_URL**[8] and **OBS_LOG**.

Any other administrative information pertaining to the file should also be included at the **INFO_URL**.

---

[7] In order to differentiate between very different factors influencing a processing step, **PRREFnx** *may* be used, where **x** is a letter A-Z. E.g., **PRREF1A** may list files and **PRREF1B** may list names of people.

[8] E.g., with **INFO_URL='http://some.site/this/guide.html'**, documents **http://some.site/this/manual.pdf** and **http://some.site/this/subdirectory/auxiliary.dat** might be harvested if it is (recursively) referenced from guide.html, but not **http://some.site/other/use.pdf**.

Proprietary data should be marked by setting the keyword **RELEASE** to the date at which the data can be freely distributed. The keyword **RELEASEC** may be used to give contact information for one or more people (name/email addresses, comma separated) administering the release details.

# *10. Reporting of events detected by the pipeline/spacecraft*

If the pipeline uses event/feature detection algorithms that will only work on the raw data, not the final pipeline product, detected events/features should be reported in pixel lists (see Appendix I-d). If possible, events that are detected during acquisition of the data but are not detectable in the acquired data should also be reported (e.g., on-board-detected events in spacecraft).

When possible, such events/features should also be reported to relevant registries following the appropriate standards (e.g., VOEvents).

# *Appendix I. Variable-keyword mechanism*

In many cases, auxiliary data such as detector temperatures, atmospheric conditions, variable exposure times, or adaptive optics performance is recorded alongside the observations. In some cases, other kinds of information such as the instrument response as a function of wavelength or a collection of instrument temperatures may be significant for correct interpretation of the data. In these cases, the variable-keyword mechanism described below can be used to link the observational data and the auxiliary data together. It can also be used to define array-valued keywords.

Since this mechanism may be used by any HDU with a non-zero **SOLARNET** keyword, we will from now on simply use the term “referring HDU” for an HDU that uses this mechanism. The actual values of a variable keyword are most commonly stored in a binary table column (called “value columns” in the description below), but image extensions may also be used (see Appendix I-d). A specification of binary table extensions can be found in Cotton et al. 1995.

To use this mechanism, the referring HDU must contain the keyword **VAR_KEYS** declaring the **EXTNAME** of the binary table extension containing the value followed by a semicolon, and then a comma-separated list of the variable keywords.

When multiple extensions are used for storing variable keyword values, this is signalled by a comma behind the last keyword of one extension, then a new **EXTNAME** followed by a semicolon, then a comma-separated list of keywords stored in that extension. The **EXTNAME** is freely chosen as long as it adheres to the **EXTNAME** rules given in Section 2.

Each keyword name may be followed by a “tag” – a string surrounded by square brackets. The tag string itself may not contain semicolons, commas, or square brackets. The tag’s only function is to distinguish between different value columns containing values for the same keyword but applicable to different referring HDUs, if such a distinction is necessary. The tag can be chosen freely – but it may be useful for humans if it is derived from the **EXTNAME** of the referring HDU. However, multiple HDUs may refer to a single tagged value column if desired, in which case it may be useful to base the tag value on all referring HDUs’ extension names.

The value columns must have **TTYPEn** equal to the keyword name plus any tag. Column numbers ( ) do not matter in the linking of value columns to keyword names. Note: *the CONTINUE Long String Keyword Convention must not be used with* **TTYPEn**, *since this is a reserved keyword defined in the FITS standard*. This means that the tag value may have to contain a shortened version of the referring HDUs’ extension names if such a scheme is used. When appropriate, it is highly recommended that the referring HDU also contains a representative scalar value of a variable keyword, although this is not mandatory. How the representative value is chosen depends on the nature of the variable keyword, though the average of the variable values is usually the appropriate choice. Variable keywords may also have string values.

As an example, the header of a referring HDU might contain the following entries:

```
EXTNAME = 'He_I    '               / Referring HDU extension name
VAR_KEYS= 'VAR-EXT-1;KEYWD_1,KEYWD_2[He_I_He_II],VAR-EXT-2;KEYWD_3'/ Variable keywords
KEYWD_1 =                      5.2 / Representative value (average) for KEYWD_1
KEYWD_2 =                        4 / Representative value (maximum) for KEYWD_2
KEYWD_3 =                        5 / Representative value (minimum) for KEYWD_3
```

This means that the values of the variable keywords **KEYWD_1** and **KEYWD_2** are stored in two separate columns in the VAR-EXT-1 binary table extension, in columns named 'KEYWD_1' and 'KEYWD_2[He_I_He_II]', respectively. Also, the **KEYWD_3** values are stored in the VAR-EXT-2 binary table extension in a column named 'KEYWD_3'. The "tag" **[He_I_He_II]** carries no intrinsic meaning, it is simply a text used to distinguish between columns in the **VAR-EXT-1** extension storing **KEYWD_2** values for different referring HDUs, e.g., **'[He_I_He_II]'** versus **'[O_V]'**. The **VAR-EXT-1** binary table header might contain the following entries (header examples from binary tables are shown in grey in this appendix):

```
EXTNAME = 'VAR-EXT-1'            / Variable keyword binary table extension name
:
TTYPE5  = 'KEYWD_1'              / Column 5: values for KEYWD_1
TTYPE6  = 'KEYWD_2[He_I_He_II]'  / Column 6: values for KEYWD_2 for He_I & He_II
TTYPE7  = 'KEYWD_2[C_II]'        / Column 8: values for KEYWD_2 for C_II
```

The TTYPE7 entry is included only to illustrate the need for the [He_I_He_II] tag in TTYPE6.

The VAR-EXT-2 binary table extension might contain the following entries:

```
EXTNAME = 'VAR-EXT-2'            / Variable keyword binary table extension name
TTYPE1  = 'KEYWD_3'              / Column 1 contains variable KEYWD_3 values
```

There are two ways in which the values of the variable keyword data cube may be associated with the data cube in the referring HDU: association by coordinates (Appendix I-a) and pixel-to-pixel association (Appendix I-b).

The mechanism described here may also be used to store a set of values that do *not* vary as a function of any coordinate or dimension of the referring HDU. Such constant, multi-valued keywords are described in Appendix I-d.

In all the examples below, the referring HDU is an image sequence with coordinates and dimensions **[x,y,t]=[HPLN-TAN, HPLT-TAN, UTC]=[512,512,60]**, with a header containing the following entries relevant to the examples in this appendix (note the formatting of **VAR_KEYS** for readability – spaces are ignored in the interpretation of the keyword):

```
DATEREF = '2023-02-01T00:00:00'   / Time coord. zero point (time reference, mandatory)
CTYPE1  = 'HPLN-TAN'              / Coord. 1 is "solar x"
CTYPE2  = 'HPLT-TAN'              / Coord. 2 is "solar y"
CTYPE3  = 'UTC     '              / Coord. 3 is time in seconds relative to DATEREF
NAXIS1  =                     512 / Size of dimension 1
NAXIS2  =                     512 / Size of dimension 2
NAXIS3  =                      60 / Size of dimension 3
:
VAR_KEYS= 'MEASUREMENTS;        &' / Extension containing measured auxiliary values
CONTINUE  '   ATMOS_R0,         &' / ATMOS_R0 values
CONTINUE  '   TEMPS,            &' / Temperatures
```

## *Appendix I-a. Variable keywords using coordinate association*

The variable-keyword mechanism using association by coordinates is fully analogous to the matching up of two separate observations – it is their shared coordinates that describe how to align the two in space, time, wavelength etc. In general, two Obs-HDUs do not necessarily have all coordinates in common. Examples are images vs. spectral rasters, or polarimetric data vs. images. Of course, the order of the WCS coordinates in the two Obs-HDUs does not matter, and e.g., the spatial, temporal, and spectral sampling of the observations may be entirely different, and the coordinates may even be irregular.

As with two separate Obs-HDUs, when using association by coordinates for variable keywords each value column has its own set of WCS keywords defining their WCS coordinates. These coordinates specify where each value in the value column data cube is located in relation to the referring HDU's WCS coordinates.

As is the case for the alignment of e.g., images vs. spectra, the value columns do not need to specify all of the coordinates in the referring HDU (e.g., a time series of temperatures vs. a sequence of images) and may have coordinates that are not present in the referring HDU (e.g., a time series of temperatures vs. a single image in the referring HDU). Furthermore, it is the coordinate *name* that is used to establish the association, after any projection has been taken into account. E.g., **HPLN-TAN** and **HPLN-TAB** are both recognised as just **HPLN** with respect to association.

If the value column contains no other coordinates than those present in the referring HDU, and no dimensions without a coordinate, only a single keyword value is associated with any pixel in the referring HDU. This is because the association uniquely determines the position within the value column based on the position in the referring HDU.

However, if the value column contains coordinates that are not present in the referring HDU, or dimensions without an assigned coordinate, there are multiple values within the value column that apply to any given pixel in the referring HDU.

***Example 1 - Variable keywords associated by shared coordinate***

Let us assume that the atmospheric coherence length **ATMOS_R0** is recorded independently during the observations described by the example header above, with a different temporal resolution than the observations.

For each exposure in the observation series, there is a single value of **ATMOS_R0** that applies to all pixels in that exposure. Thus, the **ATMOS_R0** value column should be one-dimensional, and the only coordinate that needs to be specified is time.

The header of the corresponding binary table extension MEASUREMENTS might contain the following entries:

```
EXTNAME = 'MEASUREMENTS'          / Extension containing measured auxiliary values
DATEREF = '2018-01-01T12:00:00'   / Time coord. zero point (time reference, mandatory)
TTYPE5  = 'ATMOS_R0'              / Column 5 contains values for ATMOS_R0
1CTYP5  = 'UTC     '              / Time coordinate
TDIM5   = '(4700)  '              / Array dimensions for column 5
```

As we can see, the value of 1CTYP5 is identical to the value of **CTYPE3** in the referring HDU described above. This is the only basis for determining association of coordinates. The coordinate numbers (i=1 vs **i=3**) are irrelevant in the association, it is only the values of the **UTC** coordinate together with its zero-point **DATEREF** that matters in the matching up of the two data cubes. E.g., dimension numbers and sizes (**NAXIS3** vs TDIM5) are irrelevant. Note that **DATEREF** may very well be different between the two extensions, the times that are compared are the "sum" of **DATEREF** and the **UTC** coordinate calculated according to the standard coordinate formulas!

Now, in order to find the value of **ATMOS_R0** for a given point in the referring data cube, the time corresponding to that point must be calculated. A reverse calculation is done for the value column to locate the point where its time coordinate has the same value. Then, the **ATMOS_R0** value can be extracted from that point in the value column (using linear interpolation as specified in the FITS standard).

Note that the zero point for the time coordinate (**DATEREF**) *must* be given for both extensions when one of the specified coordinates are **UTC**, and it applies to all columns with a **UTC** coordinate. Thus, if two value columns have different starting points, the relevant **iCRVLn** and/or **jCRPXn** values must be adjusted accordingly. This issue does not arise if the keyword values are stored as separate image extensions, see Appendix I-d.

If multiple values must be associated with each image, the value column would have one or more additional dimensions. The values for a given image would then be all values in the value column with a time coordinate matching that of the image. Such a scenario might arise if e.g., multiple temperatures inside the instrument is being recorded.

***Example 2 – Variable keywords associated by multiple shared coordinates***

Of course, the referring HDU and the value column may have more than one shared coordinate. In this case, the process is entirely analogous to the situation where there is only one shared coordinate: shared coordinate 1 and shared coordinate 2 are calculated for a pixel in the referring HDU, and these coordinates are then used to look up the correct value(s) in the value column, using its definition of the same coordinates.

***Example 3 – Real-life example with multi-valued keyword associated by a single shared coordinate – with table look-up of coordinates!***

As a complex, real-life example of how this mechanism can be used, we refer to CHROMIS FITS files, which contain values of R0 that are a function of time, but also a function of (two) different subfield sizes. Thus, there are two coordinates defined for the value column: time (UTC) and subfield size (WFSSZ, not a WCS coordinate). Both coordinates must be tabulated because they vary unevenly and cannot be described as linear functions of pixel coordinates. CHROMIS FITS files can be found in the SST archive at https://dubshen.astro.su.se/sst_archive/.

## *Appendix I-b. Variable keywords using pixel-to-pixel association*

Some variable keywords encode discrete-valued properties or properties that are sampled in exact sync with the observational data. In such cases, it might be important to ensure an exact correspondence between pixels in the referring HDU's data cube and pixels in the value column's data cube, without any round-off errors in the floating-point calculations of WCS coordinates.

When standard WCS calculations are used in the association between the referring HDU and the value column, such round-off errors may interfere with any exact pixel-to-pixel correspondence, resulting in a linear interpolation of the values in the value column. I.e., if a variable keyword represents a discrete-valued property, association by coordinates may result in non-discrete values. If instead a direct pixel-to-pixel association is desirable, the variable-keyword mechanism may be used as described below.

Even for non-discrete-valued keywords it may be simpler and more illustrative to use a pixel-to-pixel association. This is typically the case for values that have been measured in sync with the observations. Another example could be values varying along one detector dimension, e.g., one value per detector row.

In order to signal such an exact pixel-to-pixel association, the `WCSNn` keyword for the value column must start with `'PIXEL-TO-PIXEL'`. In this case, *no coordinate specified for the value column will be used in the association*. Also, all dimensions of the data cube in the referring HDU must be present in the value column (in the same order). Dimensions in the referring HDU for which the variable keyword has a constant value should be collapsed into singular dimensions in the value column. Trailing dimensions may be added in order to specify variable keywords with multiple values for each pixel in the referring HDU.

***Example 4 – Variable keyword with pixel-to-pixel association***

If the **`ATMOS_R0`** values from Example 1 in Appendix I-a had been recorded in sync with the 60 images, i.e., a single **`ATMOS_R0`** value is recorded for each image, the binary table extension might instead contain the following entries:

```
EXTNAME = 'MEASUREMENTS'          / Extension name of binary table extension
WCSN5   = 'PIXEL-TO-PIXEL'        / Column 5 uses pixel-to-pixel association
TTYPE5  = 'ATMOS_R0'              / Column 5 contains values for ATMOS_R0
TDIM5   = '(1,1,60)'              / Array dimensions for column 5
```

This means that the **`ATMOS_R0`** value for any referring HDU pixel **`(x,y,t)`** is found in pixel `(1,1,t)` of the `ATMOS_R0` value column data cube.

The pixel-to-pixel association may also be used if **`ATMOS_R0`** had been recorded with a lower cadence than the images. If e.g., **`ATMOS_R0`** was recorded for every 20th image then the value found in in pixel (**`1,1,1`**) of the **`ATMOS_R0`** column data cube applies to the first 20 images, the value in pixel **`(1,1,2)`** applies to the next 20 images, etc. A total of 3 **`ATMOS_R0`** measurements would have been made during acquisition of the 60 images, thus **`TDIM5='(1,1,3)'`**.

Generically, when `WCSNn='PIXEL-TO-PIXEL'`, if the size of a dimension *j* in the variable keyword data cube is *1/N* of the corresponding dimension of the data cube of the referring HDU, the pixel index $p_{j,v}$ for the variable keyword data cube can be found from the referring HDU's data cube pixel index $p_{j,d}$ through the formula $p_{j,v} = floor((p_{j,d} - 1)/N)+1$.

If multiple values must be associated with each image, the value column would have one or more additional trailing dimensions. E.g., if two independent **`ATMOS_R0`** values were measured for each image, the value column would have **`TDIM5='(1,1,60,2)'`**. Thus, for pixel **`(1,1,2)`** in the referring HDU, the values **`(1,1,2,*)`** would apply.

## *Appendix I-c. Array-valued keywords (no association)*

It is possible to use the variable-keyword mechanism to specify keywords that are multi-valued but entirely independent of the referring HDU's WCS coordinates i.e., simply an array-valued keyword. This is achieved simply by having no shared coordinates among the referring HDU and the value column. The value column may have any number of dimensions. A one-dimensional array-valued keyword is used in SPICE FITS files to record a list of lost telemetry packets for each readout window.

## *Appendix I-d. Using image extensions instead of binary tables*

The variable-keyword mechanism may also be used with image extensions instead of binary table extensions. In this case, the **VAR_KEYS** keyword contains only extension names separated with a comma and a semicolon, ending with a semicolon. The name of the variable keyword is defined by the extension name, though the extension name may also contain a tag. I.e., **VAR_KEYS='KEYWD_1 ;, KEYWD_2[He_I_He_II]; '** declares that values for **KEYWD_1** are to be found in image extension **KEYWD_1**, and values for **KEYWD_2** are to be found in image extension **KEYWD_2[He_I_He_II]**. In all other respects, this variant of the mechanism is analogous to the specification above using binary tables, with e.g., **WCSNAMEa** starting with **'PIXEL-TO-PIXEL'** to signal that pixel-to-pixel association is used.

# Appendix II. Pixel list mechanism for flagging pixels

In some cases, it is useful to flag pixels or ranges of pixels within an Obs-HDU, or to store attributes (numbers or strings) that apply only to specific pixels or ranges of pixels (see Section 5.6.2). One example is to store the location of hot/cold pixels. Another example is to store the location and original values of pixels affected by cosmic rays/spikes. Yet another example might be to highlight or label (even with a string) specific points within the data cube – such as where a reduction algorithm has broken down.

Since the pixel list mechanism described here may be used by any HDU with a non-zero `SOLARNET` keyword, we will from now on simply use the term “referring HDU” for the HDU that uses this mechanism.

This mechanism uses a specific implementation of the pixel list FITS standard (Paper I, Section 3.2), where binary table extensions are used to store pixel indices and any attributes associated with each pixel.

The binary table extension must have `N + 1 + m` columns (or only `N + m` , see special case below), where `N` is the number of data cube dimensions in the referring HDU and `m` is the number of pixel attributes (may be zero). The first columns contain pixel indices with `TTYPEn = 'DIMENSIONk'`, where `k` is the dimension number in the referring HDU. Column number `N+1` must have `TTYPEn = 'PIXTYPE'` (unless only single pixels are flagged). Any remaining columns must have `TTYPEn` set to the name of the attached attribute contained in that column, if any. Note that each cell of the binary table may only contain a single number or a string.
A zero-valued pixel index is a *wildcard* representing all allowed pixel indices in the corresponding dimension.

The `PIXTYPE` column is used to classify each of these pixels into the categories described in *Table 1* below.

| `PIXTYPE` | Meaning |
|---|---|
| 0 | Individual pixel |
| 1 | “Lower left” pixel (closest to `(1, 1, …)`) of a line/area/(hyper-)volume |
| 2 | “Upper right” pixel (farthest from `(1, 1, …)`) of a line/area/(hyper-)volume |

*Table 1: The meaning of values in the `PIXTYPE` column. To flag individual pixels, one table row is needed to specify the pixel indices of each flagged pixel. For each pixel range to be flagged, two rows are needed: one specifying the “lower left” pixel indices and the other specifying the “upper right” pixel indices.*

As a special case, the `PIXTYPE` column may be omitted if only single pixels are flagged. I.e., if no `PIXTYPE` column is present in a pixel list, all rows should be considered to be of type 0.

To establish the connection between the referring HDU and a pixel list, the referring HDU must contain the keyword `PIXLISTS`. `PIXLISTS` must declare the `EXTNAME` of the extension containing the pixel list, followed by a semicolon, then a comma-separated list of any pixel attribute names. When multiple pixel lists are used, this is signalled by adding a comma, the `EXTNAME` of the next pixel list extension followed by a semicolon, etc. Note that even when a pixel list does not contain any attributes, a comma is needed before the `EXTNAME` of any subsequent pixel list.

The EXTNAME of pixel lists may carry a meaning within the SOLARNET framework (e.g., **LOSTPIXLIST**, see Section 5.6.2). But if a pixel list **EXTNAME** ends with a “tag” (see Appendix I), this does not change its meaning. Thus, such tags may be used to distinguish between different extensions containing pixel lists of the same type/meaning for different referring HDUs. Multiple referring HDUs may refer to the same pixel list, even if it has a tag.

As an example, in order to refer to all types of pixel lists mentioned in Section 5.6.2, the *referring HDU*'s **PIXLISTS** could contain the following:

```
PIXLISTS= 'LOSTPIXLIST;, MASKPIXLIST;,     &' / Lost and masked pixels
CONTINUE  'SATPIXLIST[He_I];ORIGINAL,      &' / He_I saturated pixels w/original values
CONTINUE  'SPIKEPIXLIST[He_I];ORIGINAL,CONFIDENCE, &' / Spike pixels for He_I
CONTINUE  'SUNSPOTS;CLASSIFICATION'     / Sunspot locations and classification
```

The pixel list name **SUNSPOTS** used above is arbitrarily chosen as an example, i.e., this **EXTNAME** does not carry any predefined meaning in a SOLARNET context.

### *Example 1 – Pixel list with attribute columns*

The header of an Obs-HDU with dimensions **[lambda,x,y] = [20,100,100]** might contain the following entry:

```
PIXLISTS= 'SPIKEPIXLIST;ORIGINAL,CONFIDENCE'   / List of spike pixels
```

This means that **SPIKEPIXLIST** is a pixel list with two attribute columns, ORIGINAL and CONFIDENCE. The header of this binary table extension might include the following entries:

```
EXTNAME  = 'SPIKEPIXLIST'        / Extension name
TTYPE1   = 'DIMENSION1'          / Col.1 is index into data cube dimension 1
TTYPE2   = 'DIMENSION2'          / Col.2 is index into data cube dimension 2
TTYPE3   = 'DIMENSION3'          / Col.3 is index into data cube dimension 3
TTYPE4   = 'PIXTYPE'             / Col.4 is the pixel type
TTYPE5   = 'ORIGINAL'            / Col.5 contains original values of listed pixels
TTYPE6   = 'CONFIDENCE'          / Col.6 contains confidence values of spike detection
TCTYP1   = 'PIXEL   '            / Indicates that col. 1 is a pixel index
TCTYP2   = 'PIXEL   '            / Indicates that col. 2 is a pixel index
TCTYP3   = 'PIXEL   '            / Indicates that col. 3 is a pixel index
TPC1_1   = 1                     / Indicates that col. 1 is a pixel index
TPC2_2   = 1                     / Indicates that col. 2 is a pixel index
TPC3_3   = 1                     / Indicates that col. 3 is a pixel index
```

Here, the presence of both TCTYPn='PIXEL' and **TPCn_na=1** for n between 1 and 3 signals that this binary table extension is a pixel list, and that columns 1 to 3 are pixel indices (see Paper I, Section 3.2). Conversely, since TCTYPn is not equal to 'PIXEL' for columns 4, 5 and 6, these columns do not contain pixel indices. The use of 'PIXEL' as a coordinate name (TCTYPn) is taken from Wells et al. (1981), Appendix A, Section III-F.

Thus, if we want to flag 3 pixels in the referring HDU data cube, and store the values ORIGINAL and CONFIDENCE for each pixel, the pixel list might contain the following table values (column headings are TTYPEn values):

| | DIMENSION1 (lambda) | DIMENSION2 (x) | DIMENSION3 (y) | PIXTYPE | ORIGINAL | CONFIDENCE |
|---|---|---|---|---|---|---|
| Row 1 (pixel #1) | 5 | 10 | 1 | 0 | 500 | 0.91 |
| Row 2 (pixel #2) | 5 | 11 | 1 | 0 | 489 | 0.91 |
| Row 3 (pixel #3) | 8 | 55 | 73 | 0 | 1405 | 0.98 |

***Example 2 – Pixel list with no attribute columns***

We could list pixels that were lost during acquisition but were later filled in with estimated values. In this case, there is no original value, thus there are no attributes to associate with the pixels. An Obs-HDU might then contain:

```
PIXLISTS= 'LOSTPIXLIST[He_I];' / EXTNAME of Binary table specifying lost pixels
```

In the pixel list binary table with EXTNAME='LOSTPIXLIST[He_I]', only 4 columns would be present (**N=3**, **m=0**) and the table values might be:

| | DIMENSION1 (lambda) | DIMENSION2 (x) | DIMENSION3 (y) | PIXTYPE |
|---|---|---|---|---|
| Row 1 (pixel #1) | 1 | 10 | 3 | 0 |
| Row 2 (pixel #2) | 2 | 10 | 3 | 0 |
| Row 3 (pixel #3) | 3 | 10 | 3 | 0 |

***Example 3 – Pixel list using wildcard indices***

Assuming the data cube comes from an aggregation of exposures (scanning) in the x direction, we want to flag three hot pixels on the detector for *all* exposures (i.e., for all x indices). This is easily done using the following table in an extension with EXTNAME='MASKPIXLIST':

| | DIMENSION1 (lambda) | DIMENSION2 (x) | DIMENSION3 (y) | PIXTYPE |
|---|---|---|---|---|
| Row 1 (pixel #1) | 3 | 0 | 5 | 0 |
| Row 2 (pixel #2) | 9 | 0 | 8 | 0 |
| Row 3 (pixel #3) | 50 | 0 | 90 | 0 |

***Example 4 – Pixel list flagging of a 2-dimensional range within a 4-dimensional data cube***

As a real-life example, consider a 4-dimensional Solar Orbiter/SPICE data cube with dimensions [**x,y,lambda,t**] = [**1,1024,1024,1**]. The data was compressed onboard the spacecraft as 64 separate [**lambda,y**] = [**32,1024**] JPEG images. A telemetry packet belonging to the third of these JPEG images was lost during downlink. As a result, the third decompressed JPEG image, i.e., the (**x,y,lambda,t**) = (**1,*,65:128,1**) pixel range of the data cube, has approximated values. There are no original values or other attributes to be stored.

In the binary table pixel list we flag this pixel range by defining the “lower left” and “upper right” pixel of the pixel range by setting the PIXTYPE value to 1 and 2 respectively (see *Table 1* on page 32):

| | DIMENSION1 (x) | DIMENSION2 (y) | DIMENSION3 (lambda) | DIMENSION4 (t) | PIXTYPE |
|---|---|---|---|---|---|

| Row 1 | 1 | 0 | 65 | 1 | 1 |
|---|---|---|---|---|---|
| Row 2 | 1 | 0 | 128 | 1 | 2 |

As in Example 3, we use a zero value as a wildcard for dimension 2, representing the range 1:1024. The same effect could have been achieved using values 1 and 1024 in row 1 and two, but this might be less readable to a human who is not familiar with the data set.

The header of the pixel list binary table extension pixel list describing the approximated pixel range would contain the values listed below (among others):

```
EXTNAME = 'APRXPIXLIST[Full LW 4:1 Focal Lossy]' / Extension name

           ------------------------------
           | Column 1 specific keywords |
           ------------------------------
TTYPE1  = 'DIMENSION1'         / Pixel indices dimension 1
TCTYP1  = 'PIXEL   '           / Indicates that column 1 contains pixel indices
TDESC1  = 'Lower Left/Upper Right pixel indices of 1 approximated Lambda-Y ima&'
CONTINUE  'ge plane ranges due to loss of compressed telemetry packets&' / Axis
CONTINUE  '' / labels for column 1
           ------------------------------
           | Column 2 specific keywords |
           ------------------------------
TTYPE2  = 'DIMENSION2'         / Pixel indices dimension 2
TCTYP2  = 'PIXEL   '           / Indicates that column 2 contains pixel indices
TDESC2  = 'Lower Left/Upper Right pixel indices of 1 approximated Lambda-Y ima&'
CONTINUE  'ge plane ranges due to loss of compressed telemetry packets&' / Axis
CONTINUE  '' / labels for column 2

           ------------------------------
           | Column 3 specific keywords |
           ------------------------------
TTYPE3  = 'DIMENSION3'         / Pixel indices dimension 3
TCTYP3  = 'PIXEL   '           / Indicates that column 3 contains pixel indices
TDESC3  = 'Lower Left/Upper Right pixel indices of 1 approximated Lambda-Y ima&'
CONTINUE  'ge plane ranges due to loss of compressed telemetry packets&' / Axis
CONTINUE  '' / labels for column 3

           ------------------------------
           | Column 4 specific keywords |
           ------------------------------
TTYPE4  = 'DIMENSION4'         / Pixel indices dimension 4
TCTYP4  = 'PIXEL   '           / Indicates that column 4 contains pixel indices
TDESC4  = 'Lower Left/Upper Right pixel indices of 1 approximated Lambda-Y ima&'
CONTINUE  'ge plane ranges due to loss of compressed telemetry packets&' / Axis
CONTINUE  '' / labels for column 4

           ------------------------------
           | Column 5 specific keywords |
           ------------------------------

TTYPE5  = 'PIXTYPE '           / Pixel type
TDESC5  = 'Pixel index types: 1 = lower left corner indices, 2 = upper right&'
CONTINUE  ' corner indices' / Axis labels for column 5
```

# *Appendix III. Meta-HDU mechanism*

Most users expect to be able to analyse at least one file at a time on a laptop, preferably with all of the data loaded into memory. Thus, at some point, files become too large for comfort when following the guidelines for what to store in a single file/single Obs-HDU in a strict sense.

An obvious solution to this problem for a file that contains multiple Obs-HDUs would be to split it into multiple files containing only a single HDU each. However, at some point this strategy will not be enough to keep file sizes reasonable. E.g., simulation data are typically split into separate files for each time step and each variable. Thus, the issue of prohibitively large files should be dealt with in a more generic way while preserving the spirit of the guidelines for what should be stored together.

We do this by providing a mechanism that allows big HDUs to be split along one dimension into smaller Constituent HDUs stored in separate files and potentially individually recorded in an SVO as separately retrievable observation units, whilst also recording metadata for the original/unsplit HDU in a Meta-HDU without duplicating the data (having `NAXIS=0` or `NAXIS=1` and `NAXIS1=1`) and showing the relationship between the Constituent HDUs.

The Meta-HDU mechanism makes it easier to follow the guidelines for what to store together in a single Obs-HDU, by interpreting them as guidelines for what to store together in a single "Meta-Obs-HDU".

All Constituent HDUs must have the same `EXTNAME`, and this will be the `EXTNAME` of the HDU that results if the Constituent HDUs are stitched together.

All Constituent HDUs must have `METADIM` set to the dimension that has been split. E.g., when splitting an array `[x,y,t]` into time chunks, `METADIM=3`. Note than an accompanying auxiliary HDU with e.g., dimensions `[t,z]` would set `METADIM=1`. Auxiliary HDUs whose data array dimensions does not contain the split dimension (e.g., flatfields) do not need to contain the `METADIM` keyword. It is allowed to have `METADIM > NAXIS` to account for lost trailing singular dimensions. E.g., if constituent HDUs have dimensions `[x,y,lambda]` and `METADIM=5`, the resulting stitched HDU will have five dimensions e.g., `[x,y,lambda,1,t]`.

When possible, Constituent HDUs should contain all keywords describing the data, though some will have different values, e.g., `DATE-BEG`, `DATE-END` and keywords that vary as a function of the split dimension, or as a function of the data itself (e.g., `DATAMAX` and `DATAMIN`).

`CRPIXj` values in Constituent HDUs must refer to the same pixel in the original/unsplit HDU in order to keep all other WCS keywords identical among all Constituent HDUs. This implies that any `DATEREF` keyword should have the same value as well.

Extensions containing tabulated coordinates may also use the Meta-HDU mechanism, but they should then have `SOLARNET=-1` (as all HDUs utilizing any of the mechanisms described in this document should have).

Pixel lists should use indices that apply to the referring Constituent HDUs (and must therefore be recalculated when the Constituent HDUs are stitched together).

The original/unsplit HDU is represented by a Meta-HDU containing a comma-separated list of files containing all Constituent HDUs (**METAFILS**), and **METADIM** set to *minus* the value in the Constituent HDUs. Also, it should contain *header keywords representing the observation's global attributes* like duration, data statistics, cadence etc. A Meta-HDU may contain keywords that are not present in the constituent HDUs.

The Meta-HDU must have a set of WCS keywords that correctly describe the coordinates of the resulting stitched data array, including any added dimensions (any number of WCS coordinates may be specified irrespective of the number of dimensions in an HDU). The **WCSAXES** keyword must be set to the number of coordinates described by the set of WCS keywords.

The **EXTNAME** of such a Meta-Obs-HDU *must* be the same as the **EXTNAME** of the constituent HDUs *with the string* '**;METAHDU**' *appended* (e.g., **'He I;METAHDU'**)

Using the keywords given above, it is now possible to reconstruct/stitch together Constituent HDUs from the files given in **METAFILS** into an ideal HDU with a correct header. It is also possible to reconstruct accompanying HDUs containing their corresponding variable keyword specifications and pixel lists (though care must be taken to adjust pixel indices!).

Although we recommend having a copy of the Meta-HDUs in each constituent file, this is not a requirement. In fact, for some pipelines, it makes sense to have Meta-HDUs only in the last file, since many of the global attributes are not known until the last constituent HDUs have been processed. When necessary, the Meta-HDU may even be in a separate file.

The Meta-HDU mechanism is not restricted to Obs-HDUs, it may be applied also by any HDU with **SOLARNET=-1**.

## *Appendix III-a. Extension to multiple split dimensions*

The mechanism above may be extended to splitting HDUs along multiple dimensions. In this case, the keywords **METADIMn** should be used to indicate which dimensions have been split. It is then possible to construct Meta-HDUs resulting from aggregating Constituent HDUs along one or more dimensions. I.e., if an original HDU with dimensions [1000, 2000, 3000] has been split into 100 files/HDUs with dimensions [100,2000,300], the Constituent HDUs would have **METADIM1=1** and **METADIM2=3**.

Now it is possible to construct one set of Partial Meta-HDUs with **METADIM1=1** and **METADIM2=-3** representing the result of stitching together Constituent HDUs along dimension 3, i.e., representing a set of HDUs with dimensions (100,2000,3000).

It is also possible to construct another set of Partial Meta-HDUs with **METADIM1=-1** and **METADIM2=3** representing the result of stitching together Constituent HDUs along dimension 1, i.e., representing a set of HDUs with dimensions (1000,2000,300).

Finally, there should be a Meta-HDU with **METADIM1=-1** and **METADIM2=-3** representing the result of stitching together all Constituent HDUs, i.e., representing the original HDU with dimensions (1000,2000,3000). This final, top-level Meta-HDU should have an **EXTNAME** ending with "**;METAHDU;METAHDU**", whereas the **EXTNAMEs** of Partial Meta-HDUs should have only one "**;METAHDU**" attached. For even higher numbers of split dimensions, the rule is to add one copy of "**;METAHDU**" per layer when going from Constituent HDU towards the top level Meta-HDU.

# *Appendix IV. Adaptation to binary table extensions*

This section outlines how to adapt the SOLARNET recommendations for data stored as columns in binary table extensions (Cotton et al.). It is written for an audience that already has experience in using binary table extensions for this purpose, so many details are deliberately left out.

For any column we consider the combination of column-specific keywords (`TTYPEn`, `TDIMn`, etc), general header keywords (**`FILENAME`**, **`CREATOR`**, etc), and the associated (column) data as a self-contained quasi-HDU, entirely analogous to the normal concept of an HDU. Thus, whenever the term HDU (as in Obs-HDU) is used elsewhere in this document, it may be taken to refer to such a quasi-HDU instead of an actual HDU.

However, for such quasi-HDUs, column-specific keywords replace general header keywords according to established standards and conventions for binary tables. E.g., for column `n`, `TDIMn` replaces **`NAXIS`** and **`NAXISj`**, `TZEROn` replaces **`BZERO`** etc. Almost all WCS keywords for image extensions have binary table column equivalents. For WCS keywords without a column-specific form, the value applies to all columns. Thus, if different values of such WCS keywords are necessary for separate columns, the data *must* be placed in separate binary table extensions.

The column-specific keyword `TTYPEn` is normally used analogously to how **`EXTNAME`** is used for image extensions, but binary table extensions must also have an **`EXTNAME`** keyword set according to the rules in Section 2.

The column-specific keywords `TVARKn` replaces **`VAR_KEYS`**, and `TPXLSn` replaces **`PIXLISTS`** (see Appendix I and Appendix I-d).

The naming conventions for column-specific keywords (starting with **`T`** and allowing for 3-digit column numbers) leaves only 4 letters to carry meaning, which easily leads to the creation of very awkward column-specific keyword names. To alleviate this problem for keywords that must have different values for different columns, the column-specific keyword `TKEYSn` is introduced, listing pairs of keyword names and values inside a string. The CONTINUE Long String Keyword Convention may of course be used to improve readability and add comments, e.g.:

```
TKEYS3   = 'OBS_HDU=1,                       &' / Contains observational data
CONTINUE   'DETECTOR=″ZUN_A_HIGHSPEED2″,   &' / Detector 2
CONTINUE   'WAVELNTH=1280                   ' / [Angstrom] Principal wavelength
```

The syntax is relatively straightforward – a comma-separated list of keyword-value pairs, with string values in *double* quotes. Spaces are ignored (except inside strings).

**Warning**: non-SOLARNET-aware FITS reading software will *not* recognize values inside `TKEYSn`. Thus, FITS standard keywords – including WCS keywords – must never be given in `TKEYSn`. If no appropriate column-specific variant is valid and different values are necessary for different columns, the columns *must* instead be stored in separate binary table extensions. Thus, `TKEYSn` should be used only for project-specific and SOLARNET-specific keywords.

# Appendix V. Other recommendations or suggestions

## Appendix V-a. File naming suggestions

Although file-naming recommendations are of little consequence to an SVO, they can be a big help for users to get an overview of files stored locally, so we will give some (hopefully helpful) suggestions below.

The file name should be given in the keyword **FILENAME** (this is useful in case the actual file name is changed). **FILENAME** is mandatory for fully SOLARNET-compliant Obs-HDUs.

We recommend that file names only contain letters **A-Z** and **a-z**, digits **0-9**, periods, underscores, and plus/minus signs. Each component of the file name should be separated with an underscore – not a minus sign. In this regard, a range may be considered a single component with a minus sign between the min and max values (such as start/end date). File name components with numerical values should be a) preceded with one or more identifying letters, and b) given in a fixed-decimal format, e.g., (00.0300). Variable-length string values should be post-fixed with underscores to a fixed length.

Another common practice has been to start the file name with the “instrument name” – although typically defined in a consistent manner only on a *per mission* or *per observatory* basis – i.e., collisions may appear with other missions. Thus, we recommend prefixing the instrument name with a mission or observatory identifier (e.g., **iris** for IRIS or **sst** for SST).

After the instrument name, the data level is normally encoded as e.g., “l0” and “l1” for level 0 and 1. Note, however, that the definitions of data levels are normally entirely project/instrument-specific and does not by itself uniquely identify what kinds of processing have been applied.

Within each data set it is often very useful to have file names that can be sorted by time when subject to a lexical sort (such as with “**ls**”). This requires that the next item in the file name should be the date and time (e.g., **YYYYMMDD_HHMMSS[.ddd]** or **YYYYMMDDTHHMMSS[.ddd]**). The “**d**” part is fractional seconds, with enough digits to distinguish between any two consecutive observations.

If the data might be made available (simultaneously) with e.g., different processing emphasis (e.g., trade-offs between resolution and noise level), an alphanumeric identifier[9] of the processing mode should be added in order to ensure uniqueness of the file name.

What comes next is highly instrument-specific, but attributes that specify the type of content should definitely be encoded, e.g., which filter and wavelength has been used, which type of optical set-up has been used, etc. In particular, the wavelength is very useful for those who are not familiar with a data set.

Note that if file names (including file type suffixes) are longer than 68 characters, it will have to be split over two or more lines in a FITS header using the OGIP 1.0 long string convention. Likewise, if file name lengths exceed 67 characters, a comma-separated list of file names cannot be represented with one line per file. Thus, file names using 67 or fewer characters is preferable for human readability of FITS headers.

[9] E.g., a short form of the contents of **PRMODEn**, see Section 8.2.

The names of files containing different observations *must* be unique – i.e., all files must be able to coexist in a single directory. If the above recommendations do not result in a unique name, some additional information *must* be added. We previously recommended that file names be kept identical even if newer versions were produced, but the recommendation is now the opposite. Instead, the keyword `FILEVERP` (file version pattern) should be specified to highlight a version identifier included in the file name. E.g., a file named `solo_SPICE_sit-V01-2395.fits` and the subsequent version `solo_SPICE_sit-V02-2395.fits` should both have `FILEVERP='solo_SPICE_sit-V*-2395.fits'`, where the asterisk identifies where the file version occurs in the file name. Using an asterisk means that the file version should be interpreted lexically, whereas a percentage sign should be used when the version number is not a fixed number of decimals. E.g., with file names `file_V2.fits` and `file_V12.fits`, using `FILEVERP='file_V%.fits'` would ensure that the second file recognized as having a higher version number (thus superseding the first file).

## Appendix V-b. Storing data in a single file or in separate files

From an SVO point of view, *each Obs-HDU represents a single "observation unit"* that may be registered separately. Thus, the choice of putting such observation units into separate files or not does not matter to an SVO in terms of searchability, but it does matter in terms of the file sizes of data that may be served. However, the meta-observation mechanism may be used to circumvent this issue (see Appendix III).

However, some "typical use" aspects should still be considered – including how most existing utilities[10] interact with observations of a particular type:

As a general rule, Obs-HDUs that would typically be analysed/used together and are seldom used as stand-alone products should be stored in the same file, whereas Obs-HDUs that are often analysed/used as stand-alone products should be stored in separate files. Furthermore, Obs-HDUs with data of fundamentally different types (e.g., filter images vs. spectra vs. Fabry-Pérot data vs. spectropolarimetry) should *not* be put in the same file.

Obviously, data processed by separate pipelines cannot be stored in a single file (unless they are combined at a later stage).

**Examples and arguments in favour of a single file:**

- Data from different Stokes parameters (for the same wavelength) are normally analysed together and should be put together in a single file. In fact, different Stokes parameter values are normally stored in a single extension, with different parameters located at different data cube indices along a Stokes dimension, see Section 5.4.1. A corresponding **CTYPEi = 'STOKES'** coordinate is defined to indicate which data cube index contains data for a particular Stokes parameter. Note, however, that the Meta-HDU mechanism may be used to stitch together HDUs along the Stokes dimension so different Stokes parameters may be stored in different files.

[10] Some utilities may prefer different grouping of HDUs with respect to separate vs. single files, but that issue may be solved by a utility program that is able to join HDUs in separate files into a single file and vice versa.

- Data from a spectrometer raster[11] are normally stored in a single file, even though the data may contain information from multiple detector readout windows. They have normally been acquired in a synchronous fashion, and they may be analysed together in order to have better estimates of continuum values when performing line fitting.

- Pointing adjustments *in order to track* solar features by means of solar rotation compensation or feature tracking should not automatically cause the data to be stored in separate files. This is somewhat dependent upon the frequency and magnitude of the tracking movements relative to the field of view, and the magnitude of any time gaps relative to the cadence, but we leave it up to the discretion of pipeline designers to determine when it is appropriate to split image sequences in such cases: if the data are suitable for making a single movie, store them in a single file (and a single HDU).

- Having too many files can lead to inefficiency both on a file system basis and on the level of utilities.

- Not least, having too many files will also be an inconvenience to users who want to look at file lists manually.

**Examples and arguments in favour of separate files:**

- Observations with different **POINT_ID** values (see Section 7) should not be stored in the same file.

- Images/movies in different filters are often used as stand-alone products, even if a parallel observation in another filter exists. Thus, observations from different filters should be put in separate files.

- In a similar fashion, Fabry-Pérot scans of *separate wavelength regions* should go in separate files. The same applies to similar observations such as from spectropolarimetry.

- Some observation series with very low cadence should be stored with each image in a separate file. The definition of “very low cadence”, however, is somewhat dependent on the type of data, the resolution, and the variability time scale of resolved features. The “normal use” of the data also matters: If images are largely used as stationary context for other observations, they should definitely go into separate files. This is typically the case for synoptic observation series, which are also normally of indefinite length and therefore must be split up one way or another anyway.

[11] Rasters are observations (usually spectrometric) collected by stepping a slit across the observation area.

- Observations with significantly different starting and/or end times should *not* be stored in a single file. “Significantly different” in this context means on the order of a few times the cadence/exposure time or larger – since there may be technical reasons for differences smaller than this.

An additional aspect is that grouping data into a single file makes it impossible to download “only the interesting part” of a data set for a given analysis purpose. However, given the guidelines above, we think this is unlikely to be an issue. Also, a future SVO could provide file splitting “services”, such as selecting only specific HDUs from multi-HDU files.

When an Obs-HDU (partially) overlaps in time and space with one or more HDUs stored in other files, `CCURRENT` (concurrent) may be set to a comma-separated list of its own file name plus the names of all files containing concurrent Obs-HDUs[12].

`CCURRENT` serves as a pointer to other concurrent (and probably relevant) observations, but it also serves a purpose in grouping search results (see Section 7).

## *Appendix V-c. Obs-HDU content guidelines*

In addition to guidelines determining how data should be stored in single vs. multiple files, we here give guidelines for what should be considered as single vs. multiple observational units – i.e., *what should be stored in a single vs. multiple Obs-HDUs within each file*.

Such guidelines can only be given heuristically, due to the large diversity of possible data sets.

Each Obs-HDU will be registered as an individual observation unit, with attributes such as duration, minimum/maximum wavelength etc. Such attributes may be important search terms, and this must be taken into account when considering what should be collected into a single Obs-HDU or not.

As with the guidelines for keeping data in single or multiple files, some “typical use” aspects must also be considered – including how most existing utilities[13] interact with observations of a particular type. In addition, any “user convenience” issues should be taken into account.

**Some examples and arguments in favour of a single HDU:**

- If a Fabry-Pérot scan is stored with each exposure (each wavelength) as a separate HDU, a search made for data covering a particular wavelength inside the scan’s min/max range will not match any of the HDUs if the desired wavelength falls between the band passes of the individual exposures.

- Different Stokes parameters are so intrinsically tied to each other that they should be stored together in a single HDU. There is also an existing convention for how to do this, see Section 8.5 in the FITS Standard.

---

[12] This is of course on a “best effort” basis for the pipelines!

[13] Some utilities may prefer different practices, but a relatively simple program that is able to split or join HDUs in specific dimensions would solve the problem.

- If an observation is repeated with a more or less fixed cadence (except for small cadence variations caused by e.g., technical issues/limitations), this will not be immediately apparent if each repetition is stored as a separate HDU.

- Observations are often visualised by displaying slices of a multi-dimensional data array, sometimes also scanning through one of the dimensions in order to visualise it as a movie (though not necessarily in the time dimension). Data that are likely to be visualised in such ways should be put into a single HDU. In other words, all dimensions through which slicing or scanning may be desirable should be included in a single HDU[14].

- Pointing changes in e.g., slit-jaw movies due to slit movements should not cause the movie to be split into separate HDUs, even if there are relatively large pointing changes associated with the starting of new rasters in a series.

- Uneven spatial sampling, e.g., dense in the centre and sparse in the periphery, should not cause the data to be separated into multiple HDUs, though note that a table lookup form of WCS coordinates must be used.

- Variable exposure times due to *Automatic Exposure Control (AEC)* should not cause the exposures to be stored in separate HDUs – as long as the settings for the AEC is constant.

**Examples and arguments in favour of separate HDUs:**

- Observation units stored in separate files according to the guidelines Appendix V-b are of course stored as separate HDUs.

- If the readout of a spectrometer has gaps (i.e., only small portions of the spectrum are extracted, in “wavelength/readout windows”), the different wavelength windows should in general *not* be stored in a single HDU, since that would falsely indicate to an inexperienced user that the observation unit covers the entire spectrum between the minimum and maximum wavelengths. (It is of course *possible* to store such readout windows in a single HDU, but then one would need to use a tabulated coordinate along the “gap dimension”).

- Some observation series are made with alternating long and short exposure times. These should *not* be collected in a single Obs-HDU, because of the resulting difficulty in describing the exposure time[15] as well as the complexity that would be required in utilities in order to handle/display such data correctly. Instead, the data

---

[14] In particular, the time dimension should be included when observations are repeated and are suitable to be presented as a movie. Repeated rasters have traditionally not been collected into single HDUs, but that may be because of their low cadence - causing relatively few files to be created during a single observation run. This is changing, however, and we recommend that repeated rasters should be joined into single HDUs which include the time dimension, e.g., **`(x,y,lambda,t)`**.

[15] The variable-keyword mechanism *could* be used for e.g., **`XPOSURE`** if such data are stored in a single HDU, e.g., **`(x,y,t,2)`**, with **`XPOSURE`** stored as **`(1,1,1,2)`**. But this is much less self-explanatory than having two separate HDUs, and it would require users/utilities to be aware of the mechanisms and how to use them. Since it is not absolutely necessary to do it this way, the variable-keyword mechanism should not be used.

should be separated into one HDU with long exposures and one HDU with short exposures.

- Data that are often displayed side by side, such as images in different filters should be split into separate HDUs.

For very closely connected, parallel observations, it is preferable to handle the grouping of data into Obs-HDUs in the same way for all of the observations, even if they are not of the same type (e.g., repeated rasters and corresponding slit-jaw movies).

Bearing all of the above in mind, observations that fit the following description should be collected into a single HDU:

An array of data that has (quasi-)uniform spacing in each physical dimension, e.g., x, y, lambda, and time, and also has (quasi-)constant attributes such as pointing, exposure times, gain, filter, and other relevant settings.

# *Appendix VI. Extended mechanism for distortion corrections*

Paper V describes a mechanism for applying coordinate distortion corrections using the keyword pairs **CPDISja** and **DPja**, or **CQDISia** and **DQia**. Here we discuss the '**Lookup**' form of the mechanism, but the results may also be applied to the other forms described in Paper V.

Below we use the term "distortion array", not the equivalent "distortion data cube". This is simply to more easily distinguish between "the data cube whose coordinates are to be corrected" versus the "data array containing the distortions".

When reading Paper V as a background for this Appendix, the term "axis" should in most places be replaced by "coordinate" to match the discussion below.

With the **CPDISja**/**DPja** formulation from Paper V, the data cube's pixel coordinates are associated with the distortion array's intermediate world coordinates as a basis for lookup/interpolation of distortion values in the distortion array (Paper V, Section 3.4, last paragraph on page 9). The distortion values are then applied back to the data cube's pixel coordinates.

With the **CQDISia**/**DQia** formulation from Paper V, it is the data cube's *intermediate* pixel coordinates that are associated with the distortion array's intermediate world coordinates, and the distortion is then applied back to the data cube's *intermediate* pixel coordinates.

Thus, the original mechanism may only be used to specify coordinate distortions *based on and applied to the same coordinate stage:* pixel coordinate distortions can only be specified based on pixel coordinates, and intermediate pixel coordinate distortions can only be specified based on intermediate pixel coordinates. No other distortions are allowed, i.e., distortions can never be expressed in the final world coordinate units, not even based on the world coordinates themselves.

We have therefore chosen to extend the mechanism described in Paper V such that distortions may be specified more generally. With the extended mechanism described here, cavity errors can be represented using a map of $\Delta\lambda$ as a function of pixel coordinates, independent of tuning parameters.

We do this by introducing the keyword pair **CWDISia**/**DWia** in analogy with the **CPDISja**/**DPja** and **CQDISia**/**DQia** keyword pairs (and analogous forms for the binary table forms). Additionally, **CWERRia** replaces **CPERRja** and **CQERRia**.

As with **DPja** and **DQia** in Paper V, **DWia** is a record-valued keyword which may have multiple values in the same header for the same **i**, e.g., both **DW1='NAXES: 1'** and **DW1='AXIS.1: 1'** may occur in the same header. In text, this is written as **DW1•NAXES=1** and **DW1•AXIS.1=1** . The **DWia** records describe the association of coordinates between the distortion array and the data cube whose coordinates are to be corrected.

The specification for **CWDISia/DWia/CWERRia** given here is identical to the specification for **CPDISja/DPja/CPERRja** and **CQDISia/DQia/CQERRia** given in Paper V, except for the addition of two new record values:

**DWia•ASSOCIATE** specifies the data cube coordinate stage from which coordinate values are taken for association with the distortion array coordinates. I.e., if **DWia•ASSOCIATE** is equal to A, then data cube coordinates from stage A are used for lookup/interpolation into the distortion array to find the correction.

`DWia•APPLY` specifies the coordinate stage to which the distortion should be applied (added). I.e., if `DWia•APPLY` is equal to B, the distortion values should be added immediately after calculating the values in stage B.

All `DWia` records for a given coordinate correction should be given as contiguous sequence in the FITS header. The `DWia•APPLY` record must be the last record in this sequence.

Since record-valued keywords can only have numeric values, we must assign numbers to the coordinate stages in order to refer to them. We specify the stage numbers using a more detailed version of the normal FITS coordinate calculation flowchart, with the stage numbers given in the rightmost column:

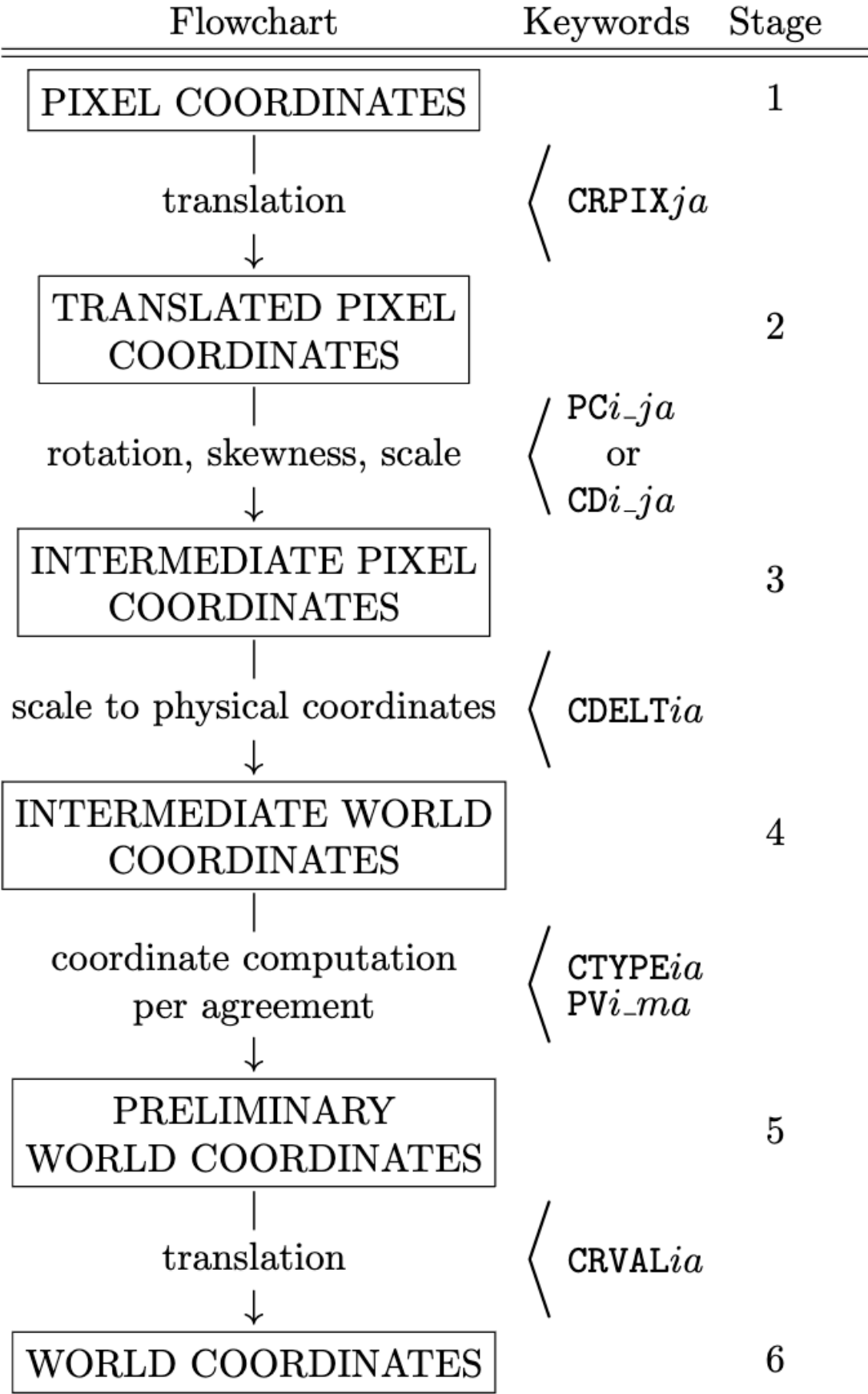

The association of coordinates between the data cube and the distortion array may happen at any of the calculation stages 1-6 for the data cube. The distortion found by interpolation in the distortion array may then be applied to the data cube coordinates at the same stage or any of the later stages. I.e., `DWia•APPLY >= DWia•ASSOCIATE`.

Thus, using the original `CPDISja/DPja` formulation corresponds to `DWia•ASSOCIATE = DWia•APPLY = 1`, and using the `CQDISia/DQia` formulation is equivalent to, and `DWia•ASSOCIATE = DWia•APPLY = 3`.

A simple example is the Solar Orbiter/SPICE pipeline (see Haugan and Fredvik 2023). After a significant re-pointing or a wheel off-loading it may take tens of minutes, even hours, until the pointing of Solar Orbiter is stable. If SPICE is observing during a period of unstable spacecraft pointing it may be necessary to take the pointing instability into account when calculating the Solar X and Solar Y coordinates of the observation. These coordinate distortion corrections are specified using `DWia•ASSOCIATE=1` and `DWia•APPLY=6`. The solar coordinates may therefore be calculated the regular way using the equation on page 8, and then the coordinate distortion corrections are applied to the world coordinates.

A more complex example, for which the original mechanism described in Paper V would be untenable, is for scanning spectrometers (e.g., Fabry-Perot),

With uneven spectrometer tuning steps, the wavelength coordinate must be tabulated using the lookup method described in Paper III. This results in a varying and *discontinuous* derivative of the translation from pixel coordinates to wavelength coordinates. Since cavity errors are measured as $\Delta\lambda$, a reverse lookup would have to be done to calculate the correct pixel or intermediate pixel coordinate distortions. If there are variations in the tuning steps over time, and/or rotations/movements of the FOV placement on the detector that are compensated for, this calculation could become prohibitive for high-speed processing of data, and it could significantly increase the file size. It would also be difficult to interpret the distortion arrays directly as distortion maps – they would be expressed in pixel or intermediate pixel coordinates, not in $\Delta\lambda$. To inspect the applied correction as a $\Delta\lambda$ distortion map, two coordinate calculations would have to be performed – one with corrections and one without corrections – to see the difference between the two.

In the SSTRED pipeline (see Löfdahl et al. 2021), cavity errors are specified using `DWia•ASSOCIATE=1` and `DWia•APPLY=6`. Thus, cavity errors are specified by lookup/interpolation using the data cube's pixel coordinate system but applied to the world coordinates.

Paper V implicitly states that the distortion array's intermediate world coordinates should be used during the association with the data cube's coordinates – i.e., after application of the distortion array's `CRPIXj`/`CDELTi`/`CRVALi`, but they exclude the use of `PCi_j`. In other words, the distortion array coordinates may not be rotated nor sheared, but must rather be defined on a rectangular, equidistant grid. We see no reason to keep this restriction, but rather define the association to be performed *using the distortion array's world coordinates, however they may be specified*. Thus, specifications of coordinates for the distortion array follow the same rules as for any other HDU, and it is the final coordinates of the distortion array that should be used during the association of coordinates between the distortion array and the data cube.

This will allow specification of distortion arrays with rotated and, not least, non-equidistant coordinates, which may make more sense for some applications, and may also save significant space.

As mentioned initially, the extension of Paper V's formalism may also be applied to distortion specifications other than `'Lookup'`. I.e., if `DWia•ASSOCIATE` is equal to A, then data cube coordinates from stage A are used as arguments to the specified distortion function in order to calculate the distortion values.

The resulting distortion values are then applied (added) to the coordinate values specified by `DWia•APPLY`. I.e., if `DWia•APPLY` is equal to B, the distortion values should be added immediately after calculating the values in stage B.

## *Appendix VI-a. Multiple distortion corrections to the same coordinate*

If e.g., one distortion is best specified using a polynomial and another distortion of the same coordinate must be specified by table lookup, they can be specified individually using multiple corrections. Multiple distortion corrections are specified by multiple sets of **DWia** records, each ending with **DWia•APPLY**. No **DWia** record applying to a subsequent correction may occur before the **DWia•APPLY** record of a previous correction.

**DWia•APPLY** values may never decrease from one correction to the next. When using multiple corrections for a single coordinate, they should be separated by a **COMMENT** line, to decrease the theoretical chance that some FITS processing software reorders the keywords.

Distortions with the same **DWia•APPLY** values are added together.

If a subsequent correction has **DWia•ASSOCIATE** equal to the **DWia•APPLY** of a previous correction, it is the *uncorrected coordinate values* from the **DWia•ASSOCIATE** stage that should be used as a basis for the subsequent correction.

If instead a subsequent correction must be performed on *corrected coordinate values* from a previous correction, chaining of corrections may be specified using fractional values for the **DWia•APPLY** and **DWia•ASSOCIATE** records. Fractional **DWia•APPLY** values implicitly define new coordinate stages in between the integer-valued stages defined here, which may then be referenced by a subsequent **DWia•ASSOCIATE** record.

E.g., if one correction has **DWia•ASSOCIATE=1** and **DWia•APPLY=1.1**, a subsequent correction may have **DWia•ASSOCIATE=1.1**, meaning that the coordinate values *after* the first correction should be used for association during the second correction.

# *Appendix VII. External Extension References*

In SOLARNET FITS files, in keywords that have no special function within the official FITS/WCS framework, references to other extensions may be in the form of *external extension references*, e.g.:

```
VAR_KEYS= '../auxiliary/s35837r001-aux.fits;VAR_KEY_DATA;TEMPERATURE[He_I]'
```

This means that the file with a relative path of `'../auxiliary/'` and a file name of `'s35837r001-aux.fits'` contains a binary table with `EXTNAME='VAR_KEY_DATA'`, containing a column with `TTYPEn='TEMPERATURE[He_I]'` which holds the data for the variable keyword `TEMPERATURE`. The path to the referenced file (`s35837r001-aux.fits`) is relative to the path of the referencing file.

In general terms, external extension references have the form `'<relative-path>/<filename>;<extname>'` and are drop-in replacements wherever plain extension names may be used. The relative path must always start with either `'./'` or `'../'`. In the example above, the external extension reference is followed by a column name, just as if the external extension reference was a regular extension name. When looking up the file to locate the referenced extension, the software should allow for any file name variations due to compression (e.g., endings like .gz and .zip).

**Placeholder extensions**

Of course, it may be that the end user does not have the file containing the external extension available. To partially amend this situation, it is possible to have a placeholder extension in the same file as the referring extension, containing the full header of the referenced extension but having only a degenerate data cube (i.e., `NAXISn=1`). The `EXTNAME` of the placeholder extension must be identical to the `EXTNAME` of the true external extension (i.e., `'VAR_KEY_DATA'` in the example above). The data dimensions of the true external extension should then be given in `XDIMNAn=NAXISn` of the true external extension.

# Appendix VIII. File list glob patterns and sorting

In SOLARNET fits files, keywords used to give comma-separated lists of files may use the shell glob patterns asterisk (*) matching any number of characters, question mark (?) matching a single character, and character set (e.g., `[ABCx-y]`) matching a single character as specified within the brackets. The files matching the pattern should be sorted in lexicographic order before being interpreted as a list of file names. File lists may also use the relative path notation as specified in Appendix VII.

# *Appendix IX. Higher-level data: parameterized components*

One common type of higher-level data are results from analysing lower-level data by fitting of parameterized components (e.g., emission line profiles) to spectroscopic data by means of $\chi^2$ minimization, but so far there has been no standard mechanism for how to store such results in FITS files.

Below we describe a recommended scheme for storing such results, comprehensive enough to store any data resulting from fitting of additive and multiplicative parameterized components. The scheme allows for later manual inspection, verification, and (if desirable) modification of the results. We will refer to files using this scheme as "(SOLARNET) Type P". We suggest that "P" is used as a suffix to the relevant data level number for such data. E.g., Solar Orbiter SPICE files using this scheme are referred to as SPICE Level 3P. *These files should be considered as a reference implementation of this recommendation* and will be used as an example below. Below we use dimensions `[x,y,lambda,t]` simply as an example, since those are the dimensions used in SPICE Level 3P FITS files.

For a typical SPICE Level 2 data cube with dimensions `[x,y,lambda,t] = [400,400,32,100]`, fitting of a single Gaussian plus a zero-order polynomial is made for every `(x,y,t)` position. The final result is a data cube `[x,y,t,p] = [400,400,100,5]` where

- `(x,y,t,1)` is the fitted line peak intensity $I_0$
- `(x,y,t,2)` is the fitted line center $\lambda_c$
- `(x,y,t,3)` is the fitted line width $w$
- `(x,y,t,4)` is the fitted constant background $a_0$ (in a zeroth-order polynomial)
- `(x,y,t,5)` is the $\chi^2$ value from the fit

Thus, such SPICE Level 3P data are the best fitting parameters $(\lambda; I_0, \lambda_c, w, a_0)$ for the function:

$$F(\lambda; I_0, \lambda_p, w, a_0) = Gaussian(\lambda; I_0, \lambda_c, w) + Polynomial(\lambda; a_0)$$

for each point `(x,y,t)`.

For readout windows with multiple significant emission lines, multiple Gaussians are used. When e.g., two Gaussians are used, the Level 3P data will be the best-fitting parameters $(I_{0_1}, \lambda_{p_1}, w_1, I_{0_2}, \lambda_{p_2}, w_2, a_0)$ of the function:

$$F(\lambda; I_{0_1}, \lambda_{c_1}, w_1, I_{0_2}, \lambda_{c_2}, w_2, a_0) = Gaussian(\lambda; I_{0_1}, \lambda_{c_1}, w_1) + Gaussian(\lambda; I_{0_2}, \lambda_{c_2}, w_2) + Polynomial(\lambda\,; a_0)$$

for every point `(x,y,t)`, giving a Level 3P data cube with dimensions `[x,y,t,p] = [400,400,200,8]`, where `(x,y,t,1…3)` is $(I_{0_1}, \lambda_{p_1}, w_1)$, `(x,y,t,4..6)` is $(I_{0_2}, \lambda_{p_2}, w_2)$, `(x,y,t,7)` is $a_0$, and `(x,y,t,8)` is the $\chi^2$ value from the fit.

Generally, for *n* Gaussians and a constant background, the size of the parameter dimension would be 3n+1+1. For n Gaussians and a linear background, the size would be 3n+2+1 because the last component would be $Polynomial(\lambda; a_0, a_1) = a_0 + a_1\lambda$. Additional components may be defined, e.g., Voigt profiles and instrument-specific components (broadened Gauss profiles for SOHO/CDS).

Since the lambda coordinates for `(x,y,*,t)` are passed to the fitting function together with the corresponding data points, we refer to the lambda dimension as a "fitting dimension", whereas dimensions `x`, `y`, and `t` are referred to as "result dimensions". In principle, both the fitting dimensions and the result dimensions may be entirely different and in a completely different order for other types of data (e.g., a `STOKES` dimension may be included).

Since the lambda dimension does not appear in the resulting data cube, it is said to be "absorbed" by the fitting process.

In general, the scheme can be used to store data analysis results from fitting of any function on the form:

$$F(\lambda;\boldsymbol{p}) = \Big( ( f_1(\lambda;\boldsymbol{p}_1) + f_2(\lambda;\boldsymbol{p}_2) + \cdots ) \cdot f_j\big(\lambda;\boldsymbol{p}_j\big) + \cdots \Big) \cdot f_x(\lambda;\boldsymbol{p}_x) + \cdots$$

where $f_n$ are individual components, $\boldsymbol{p}_n$ are their parameters, and $\boldsymbol{p}$ is the aggregation of all parameters, by $\chi^2$minimization of:

$$\chi^2 = \sum_{\lambda} W(\lambda) \cdot \big(y(\lambda) - F(\lambda;\boldsymbol{p})\big)^2$$

where $W(\lambda)$ is the statistical weight of each pixel (typically $1/\sigma^2$) and $y(\lambda)$ is the original data. The bold font for $\boldsymbol{p}$ and $\boldsymbol{p}_n$ indicates vectors of parameters, distinguishing them from individual parameters in non-bold font.

To ensure that the result of the analysis can be interpreted correctly, the full definition of $F(\lambda;\boldsymbol{p})$ and its parameters must be specified in the header of the extension containing the result, using the following keywords:

### Mandatory general keywords for HDUs with SOLARNET Type P data

`SOLARNET` must be set to either `0.5` or `1`, and `OBS_HDU=2` (*not* `1`*!*) signals that the HDU contains SOLARNET Type P data.

`ANA_NCMP` must be set to the number of components used in the analysis.

The `CTYPEi` of the parameter dimension must be `'PARAMETER'`. Note that the Meta-HDU mechanism (Appendix III) may be used to split Type P data over multiple files along this dimension, so e.g., parameters from each component are stored in separate files. In such cases, all HDUs should contain a full complement of all keywords defined here (including those describing components whose parameters are not present in the file).

### Mandatory keywords describing each component

`CMP_NPn`: Number of parameters for component `n`

`CMPTYPn`: Component type, e.g.,

- `'Polynomial'`, a polynomial $p_1 + p_2\lambda + p_3\lambda^2 + \cdots$ of order `CMP_NPn - 1`.
- `'Gaussian'`, a Gaussian $p_1 e^{-\frac{1}{2}(\lambda-p_2)^2/p_3^2}$
- `'SSW comp_bgauss'`, a broadened Gaussian $f(p_1,\cdots,p_5)$, see SSW routine `comp_bgauss`
- `'SSW comp_voigt'`, a Voigt profile $f(p_1,\cdots,p_4)$, see SSW routine `comp_voigt`
- … etc. If you need additional component types, contact prits-group@astro.uio.no.

### Optional *functional* keywords for each component

`CMPMULn`: Indicates whether component `n` is multiplicative (`CMPMULn=1`), in which case it is to be applied to the result of all previous components `n-1`, `n-2`, etc. This can be used for e.g., extinction functions. Default value is `0`.

`CMPINCn`: Indicates whether component `n` is included (`CMPINCn=1`) or excluded (`CMPINCn=0`) in the fit. This allows specification of components that would normally be included but for some reason (e.g., poor S/N ratio) has been left out for this particular data set. If `CMPINCn` is zero, additive components have a value of zero independent of the parameter values, and multiplicative components have a value of 1. Default value is 1, i.e., the component is included in the fit. The parameters in the data cube should be set to the initial values that would have been used if the component was included. If this is not feasible, the parameters should be set such that the component value would be zero if it had in fact been included.

### Optional *descriptive* keywords for each component

`CMPNAMn`: Name of component `n`, typically used to identify/label the emission line fitted, e.g., `'AutoGauss79.5', 'He I 584'.`
`CMPDESn`: Description of component `n`
`CMPSTRn`: Used by SSW's `cfit` routines for the string to be included in composite function names. E.g., the function string for a Gaussian is '`g`', and '`p1`' for a first-order polynomial. An automatically generated composite function of the two would be called '`cf_g_p1_`'.

Each component can have up to 26 parameters labelled with `a=A-Z`, corresponding to component parameters $p_\mathrm{a} = p_1 \ldots p_{26}$.

### Mandatory functional keyword for each parameter

`PUNITna`: The units for parameter `a` of component `n`, specified according to the FITS Standard Section 4.3, e.g., `'nm' or 'km/s'`.

### Optional functional keywords for each parameter

`PINITna`: Initial value for $p_\mathrm{a}$ used during fitting
`PMAXna`: Maximum value ($p_\mathrm{a}$ has been clamped to be no larger than `PMAXna` during fitting)
`PMINna`: Minimum value ($p_\mathrm{a}$ has been clamped to be no less than `PMINna` during fitting)
`PCONSna`: Set to 1 if $p_\mathrm{a}$ has been kept constant during fitting
`PTRAna`: Linear transformation coefficient $A$ (default value 1), see below
`PTRBna`: Linear transformation constant $B$ (default value 0), see below

When `PCONSna=1`, i.e., when $p_\mathrm{a}$ has been kept constant during fitting, it does not mean that $p_\mathrm{a}$ necessarily has the same value for all `(x,y,t)`. The parameter may have been set to different values at different points prior to the fitting e.g., manually, and then not been allowed to change during subsequent fitting of the other parameters. For points where a parameter has been kept constant, the `PINITna` value does not apply. Using the example above, the data cube value for `(x,y,t,p)` can differ from `(x,y,t+1,p)` even if the corresponding `PCONSna` value is `1`. It is also possible to keep a parameter constant only at specific points `(x,y,t)` using a constant mask in a separate extension with the same dimensionality as the result data cube (except for the last dimension, which will be one smaller than in the result data cube because there is no constant mask for the $\chi^2$ value). If the constant mask extension is present, parameter number `p` has been kept constant/fixated for `(x,y,t)` at the value given in the result data if and only if the constant mask `(x,y,t,p)=1`. Thus, values in the constant mask overrides the `PCONSna` value.

A Gaussian component is explicitly defined to be simply $f(\lambda; p_1, p_2, p_3) = p_1 e^{-\frac{1}{2}(\lambda-p_2)^2/p_3^2}$. However, some may prefer to store results in modified form, such as velocities instead of line positions, and with varying definitions of line width (e.g., FWHM). To accommodate this without having to create separate components for every form, it is possible to use the `PTRAna` and `PTRBna` keywords to define a linear transformation between the *nominal* (stored) value $n_\mathrm{a}$ of a parameter and the *actual* value $p_\mathrm{a}$ that is passed to the component function.

Given $A_\mathrm{a}$=`PTRAna` and $B_\mathrm{a}$=`PTRBna`, the actual parameter value passed to the component function is $p_\mathrm{a} = A_\mathrm{a} \cdot n_\mathrm{a} + B_\mathrm{a}$ and conversely $n_\mathrm{a}=\frac{p_\mathrm{a}-B_\mathrm{a}}{A_\mathrm{a}}$. If we set $A = \frac{\lambda_0}{c}$ and $B = \lambda_0$ then:

$$n_a = \frac{c(p_a - \lambda_0)}{\lambda_0}$$

Thus, if $p_\mathrm{a} = \lambda_c$ (the fitted line centre) then the nominal value stored in the data cube is $n_\mathrm{a} = v$ (the line velocity, with positive values for red shifted lines).

Likewise for the third parameter of a Gaussian, if $A = \frac{1}{2\sqrt{2\ln 2}}$ then the nominal value stored in the data cube will be the full width of half maximum (FWHM).

**Optional descriptive keywords for each parameter**

`PNAMEna`: Parameter name, e.g., `'intensity'`, `'velocity'`, `'width'`
`PDESCna:` Parameter description

**Optional functional keywords for the analysis as a whole**

`PGFILENA`: the name of the (progenitor) file containing the original data
`PGEXTNAM`: the name of the extension in the progenitor file that contains the original data
`NXDIM`: the number of dimensions absorbed by the fitting process
`XDIMTYm`: the `CTYPEi` of the `m`th dimension absorbed by the fitting process.

To allow full manual inspection, verification, and modification of the analysis results, several auxiliary data arrays may be stored in separate HDUs, with their `EXTNAME` given in the following keywords. In the description we specify the dimensionalities that would result from the example discussed above.

- `RESEXT:` the HDU containing the analysis results (`[x,y,t,p]`). Note `OBS_HDU=2`
- `DATAEXT:` the original data/Obs-HDU (`[x,y,lambda,t]`)
- `WGTEXT:` data weights $w$ used during fitting (`[x,y,lambda,t]`). When not present, all data points are assumed to have equal weight.
- `RESIDEXT:` residuals from the fitting process (`[x,y,lambda,t]`) which may in some cases be an important factor in the verification e .g., to discover emission lines that have not been considered during the fitting
- `CONSTEXT:` constant mask (`[x,y,t,p]`) – if the constant mask value `(x,y,t,p)=1`, parameter `p` has been kept constant/frozen at the stored value during the fitting process for point `(x,y,t)`. When the constant mask extension is not present, it is assumed that all parameters have been fitted freely (between the specified min and max values) at all times unless **PCONSna=1**.

- **INCLEXT:** component inclusion mask (**[x,y,t,n]**) – if **(x,y,t,n)=0**, component n has not been included for point (x,y,t). When not present, it is assumed that all components have been included at all times.
- **XDIMXTm:** The values of coordinates absorbed by the fitting process (**[x,y,lambda,t]**) specified by **XDIMTYm**, (see above) may be included in separate extensions as a convenience, although this is redundant whenever any of the arrays with all dimensions of the original data is present and contains the appropriate WCS information. Thus, **XDIMXTm** refers to the extension containing absorbed coordinate number **m** (with coordinate specification given by **XDIMTYm**).

In all such extensions, all WCS keywords that apply must be present, given their dimensionalities, as must all Type P-related keywords (including e.g., the extension names and component/parameter descriptions etc., and **OBS_HDU=2** as these are also “type P” data). For the component inclusion mask extension (**INCLEXT**), the **CTYPEi** of the component dimension should be **'COMPONENT'**.

For these auxiliary extensions, it may be worth considering the “external extensions” mechanism, see Appendix VII.

At the time of writing, the SPICE Level 3P pipeline is not yet set in stone, and no Level 3P data has been delivered to the Solar Orbiter archive, thus there is no lock-in of the definitions yet. Please inform us at prits-group@astro.uio.no if you implement this mechanism.

### Extension to other types of higher-level data

The Type P storage scheme may also be used for results from other types of analyses that do not involve forward modelling of the data and subsequent $\chi^2$ minimisation, as a way to store interrelated parameters that have been determined from data in other ways, e.g. Mg II k line parameters, with **CMPNAMn='Mg II k'**, and **PNAMEna** set to e.g., **'k2v'**, **'k2r'**, or **'k3'**. For such cases, other values for the **CMPTYPn** keywords must be found (contact prits-group@astro.uio.no), and the size of the **PARAMETER** dimension will be equal to the sum of the **CMP_NPn** keywords, not the sum plus 1 as is normally the case.

# 11. References

- The FITS Standard, version 4.0: Definition of the Flexible Image Transport System (FITS), https://fits.gsfc.nasa.gov/fits_standard.html. (Published version, Pence et al, 2010, A&A, 524, A42, 40 pp.)

- Paper I: Representations of World Coordinates in FITS (Greisen & Calabretta, 2002, A&A, **395**, 1061-1075, http://www.aanda.org/articles/aa/pdf/2002/45/aah3859.pdf)

- Paper II: Representations of celestial coordinates in FITS (Calabretta & Greisen, 2002, A&A, **395**, 1077-1122, http://www.aanda.org/articles/aa/pdf/2002/45/aah3860.pdf)

- Paper III: Representations of spectral coordinates in FITS (Greisen, Calabretta, Valdes & Allen, 2006, A&A, **446**, 747-771, http://www.aanda.org/articles/aa/pdf/2006/05/aa3818-05.pdf)

  - Authors' web sites, supplemental background: Eric Greisen, Mark Calabretta, http://www.atnf.csiro.au/people/mcalabre/WCS/index.html

  - An unofficial errata for Papers I, II, and III (Calabretta & Greisen, http://fits.gsfc.nasa.gov/wcs/errata_20071222.pdf)

- Paper IV: Representations of Time Coordinates in FITS (Rots, 2015, A&A, **574**, A36, https://www.aanda.org/articles/aa/pdf/2015/02/aa24653-14.pdf)

- Paper V: Representations of distortions in FITS world coordinate systems (Calabretta, Valdes, Greisen, Allen, ADASS, 2004, **314**, http://fits.gsfc.nasa.gov/wcs/dcs_20040422.pdf)

- Coordinate systems for solar image data (Thompson, 2006, A&A, **449**. 791-803, http://www.aanda.org/articles/aa/pdf/2006/14/aa4262-05.pdf)

- FITS: A Flexible Image Transport System (Wells et al, 1981, A&AS, **44**, 363)

- Precision effects for solar image coordinates within the FITS world coordinate system (Thompson, 2010, A&A, **515**, A59, http://www.aanda.org/articles/aa/pdf/2010/07/aa10357-08.pdf).

- The SolarSoft WCS Routines: A Tutorial (Thompson, http://hesperia.gsfc.nasa.gov/ssw/gen/idl/wcs/wcs_tutorial.pdf)

- SSTRED: Data- and metadata-processing pipeline for CHROMIS and CRISP (Löfdahl et al. 2021, A&A, **653**, A68)

- Solar Orbiter SPICE Data Product Description Document (Haugan and Fredvik, 2023, https://spice-wiki.ias.u-psud.fr/doku.php/data:data_analysis_manual)

- [S-META-SIM]: SOLARNET Metadata Recommendation for Simulated Data (Haugan and Fredvik, 2023, http://sdc.uio.no/open/solarnet/ , https://arxiv.org/????????

- Binary table extension to FITS (Cotton et al., 1995, A&AS, **113**, 159-166, http://adsabs.harvard.edu/abs/1995A%26AS..113..159C).

- Checksum Keyword Convention (http://fits.gsfc.nasa.gov/registry/checksum.html)

- The FITS Header Inheritance Convention (https://fits.gsfc.nasa.gov/registry/inherit/fits_inheritance.txt

- The CONTINUE Long String Keyword Convention (https://fits.gsfc.nasa.gov/registry/continue_keyword.html)

- Space Physics Archive Search and Extract (SPASE) instrument types (http://www.spase-group.org/data/reference/spase-2_2_8/spase-2_2_8_xsd.htm - InstrumentType and http://www.spase-group.org/docs/dictionary/spase-2_2_8.pdf)
- Recommendations for Data & Software Citation in Solar Physics (2012AAS…22020127H)
- Best Practices for FITS Headers (2012AAS...22020128H, http://sdac.virtualsolar.org/docs/SPD2012/2012_SPD_FITS_headers.pdf)
- VSO Checklists, http://virtualsolar.org/checklists
- VSO Minimum Information for Solar Observations, http://docs.virtualsolar.org/wiki/MinimumInformation
- The Unified Content Descriptors, Version 1+ (UCD1+) http://www.ivoa.net/documents/latest/UCDlist.html

**Other sources of keywords with established use:**

- Metadata Definition for Solar Orbiter Science Data (https://issues.cosmos.esa.int/solarorbiterwiki/display/SOSP/Metadata+Definition+for+Solar+Orbiter+Science+Data)
- STEREO Standard FITS keywords (http://jsoc.stanford.edu/doc/keywords/STEREO/STEREO_site_standard_fits_keywords.txt)
- SDO/AIA FITS keyword definitions (https://www.lmsal.com/sdodocs/doc?cmd=dcur&proj_num=SDOD0019&file_type=pdf)
- IRIS FITS keyword definitions (http://www.lmsal.com/iris_science/irisfitskeywords.pdf)

# *Part B. Lists of mandatory and optional FITS keywords with example values*

## *12. Mandatory keyword for all HDUs (Section 2.1)*

In addition to all keywords required by the FITS Standard, all HDUs (including the primary HDU) in SOLARNET FITS files must contain the keyword **EXTNAME**, with a value that is unique within the file.

```
EXTNAME= 'He_I    '           / Name of HDU
```

## *13. Mandatory keywords for all Obs-HDUs (Section 2.2)*

```
SOLARNET=                  0.5 / Fully SOLARNET-compliant=1.0, partially=0.5
OBS_HDU =                    1 / This HDU contains observational data
DATE-BEG= '2020-12-24T17:12:00.5' / Date of start of observation
```

## *14. Mandatory WCS keyword for all HDUs with a UTC (time) coordinate (Section 4.1)*

```
DATEREF = '2020-12-24T00:00:00' / Time coordinate zero point
```

## *15. Mandatory keywords for fully SOLARNET-compliant Obs-HDUs*

The keywords listed in this section are mandatory for fully SOLARNET-compliant Obs-HDUs. However, most keywords are only “conditionally mandatory”, depending on the data content, which mechanisms have been used, instrument type, etc.

### 15.1. Mandatory general keywords (Sections 8, 9, and Appendix V-a)

```
FILENAME= 'sleep_a_zen_l2_20201224_170000.1_balanced.fits'
DATASUM = '250353142'         / Data checksum
CHECKSUM= 'hcHjjc9ghcEgh9g'   / HDU checksum
DATE    = '2020-12-31T23:59:59' / Date of FITS file creation
```

```
ORIGIN  = 'University of Oslo'  / Location where FITS file has been created
```

## 15.2. Fundamental WCS coordinate keywords (Section 3.1)

Obs-HDUs must contain all WCS coordinate specifications that are required to *adequately describe the observations* for their normal use. This includes e.g., the use of extra coordinates for singular dimensions when necessary (i.e., **WCSAXES** may be greater than **NAXIS**), or the use of alternate WCS coordinate systems (with WCS keywords ending in a letter **A-Z**). Normally, WCS keywords that must be included are **CRVALi**, **CDELTi**, **CRPIXj**, **CUNITi**, **CTYPEi**, and when necessary, also e.g., **WCSAXES**, **PCi_j** (or **CDi_j**), **CRDERi**, and **CSYERi**.

The example below is a brief excerpt of a header describing an observation where the random and systematic errors in the time coordinate are so large that they may be important for certain types of analysis.

```
CTYPE3  = 'UTC     '           / Coordinate 3 is time
CUNIT3  = 's       '           / Units for time coordinate
CRPIX3  =                    1 / Reference pixel: time starts at first image
CRVAL3  =                    0 / [s] Offset from DATEREF of reference pixel
CDELT3  =                  0.1 / [s] Sampling is 0.1 seconds
CRDER3  =                 0.03 / [s] Large random clock error (sample-to-sample)
CSYER3  =                    5 / [s] Large systematic error, clock may be off by 5s
```

## 15.3. Mandatory WCS positional keywords (Section 3.2)

### *15.3.1. Mandatory for ground-based observatories (Section 3.2)*

```
OBSGEO-X=         5327395.9638 / [m] Observer's fixed geographic X coordinate
OBSGEO-Y=        -1719170.4876 / [m] Observer's fixed geographic Y coordinate
OBSGEO-Z=          3051490.766 / [m] Observer's fixed geographic Z coordinate
```

### *15.3.2. Mandatory for Earth orbiting satellites (Section 3.2)*

```
GEOX_OBS=         1380295.0032 / [m] Observer's non-fixed geographic X coordinate
GEOY_OBS=           57345.1262 / [m] Observer's non-fixed geographic Y coordinate
GEOZ_OBS=         9887953.9454 / [m] Observer's non-fixed geographic Z coordinate
```

### *15.3.3. Mandatory for deep space missions (not Earth orbiting satellites) (Section 3.2)*

```
HGLN_OBS=           -0.0572950 / Observer's Stonyhurst heliographic longitude
HGLT_OBS=              5.09932 / Observer's Stonyhurst heliographic latitude
DSUN_OBS=        88981577950.3 / [m] Distance from instrument to Sun centre
```

## 15.4. Mandatory data description keywords (Sections 5.1, 5.2 and 5.6.2)

The **BTYPE** keyword should contain either a UCD description of the data contents, or a more human readable description[16]:

```
BTYPE   = 'phot.radiance;em.UV' / Unified Content Descriptors v1.23
```
*or*
```
BTYPE   = 'Spectral Radiance'   / Type of data
```

Note that if a UCD description is given, a human readable description may instead be given by the optional keyword **BNAME**. On the other hand, if **BTYPE** gives a human readable description, the UCD description may instead be given by the optional keyword **UCD**. See Section 18.5 for examples.

```
BUNIT   = 'W/m2/sr/nm'          / Physical units of calibrated data
XPOSURE =                  2.44 / [s] Accumulated exposure time
```

When the data are a result of multiple summed exposures with identical exposure times, the keyword **TEXPOSUR** and **NSUMEXP** must be set:

```
TEXPOSUR=                  1.22 / [s] Single-exposure time
NSUMEXP =                     2 / Number of summed exposures
```

**NBINj** and **NBIN** is mandatory if the data has been binned:

```
NBIN1   =                     2 / Binning in dimension 1
NBIN2   =                     4 / Binning in dimension 2
NBIN    =                     8 / Product of all NBINj
```

Missing or blank pixels in floating-point-valued HDUs should be set to *NaN*, but missing or blank pixels in integer-valued HDUs must be given the value of **BLANK**:

```
BLANK   =                  -100 / Value of missing pixels (integer HDU)
```

## 15.5. Mandatory keywords identifying the origin of the observations (Section 6)

A *subset* of the following keywords is mandatory in the sense that the subset must be sufficient to uniquely identify the origin of the observations, and they should be present to the extent that they make sense for the given observations (e.g., **MISSION** might not make sense for ground-based observations, or there might be no sensible value for **PROJECT**).

```
PROJECT = 'Living With a Star' / Name of project
MISSION = 'SLEEP   '           / Name of mission
OBSRVTRY= 'SLEEP A '           / Name of observatory
TELESCOP= 'ZUN     '           / Name of telescope
TELCONFG= 'STANDARD'           / Telescope configuration
INSTRUME= 'ZUN     '           / Name of instrument
CAMERA  = 'cam1    '           / Name of camera
GRATING = 'GRISM_1 '           / Name of grating/grism used
FILTER  = 'Al_med, open'       / Name of filter(s)
```

---

[16] For simulated data and simulated observations, **BTYPE** can be a plot axis description as in published Bifrost FITS files, with a further description given in **BNAME**, see Sections 5.1 and 15.1 of [S-META-SIM].

```
DETECTOR= 'ZUN_A_HIGHSPEED1'    / Name of detector
OBS_MODE= 'lo-res-hi-speed12b' / Name of predefined settings used during obs.
SETTINGS= 'fpos=123,vpos=3'     / Additional instrument/acquisition settings
OBSTITLE= 'High cadence active region raster'/ Title of observation
OBS_DESC= 'High cadence raster on AR10033'/ Description of observation
OBSERVER= 'Charlotte Sitterly' / Who acquired the data
PLANNER = 'Natalia Stepanian'  / Observation planner
REQUESTR= 'Annie Maunder'      / Who requested this particular observation
AUTHOR  = 'Cecilia Payne'      / Who designed the observation

CAMPAIGN= 'FlareHunt791,JOP922'/ Coordinated campaign name(s)
CCURRENT= 'sleep_a_zen_l2_20201224_170000_120_balanced.fits,&' / Concurrent,
CONTINUE  'sleep_b_zen_l2_20201224_170001_045_balanced.fits,&' / overlapping
CONTINUE  'sleep_a_magneto_l2_20201224_170001_030_full.fits,&' / observations
CONTINUE  'sleep_b_magneto_l2_20201224_170001_808_full.fits,&' / (multiple files)
CONTINUE  'sleep_b_magneto_l2_20201224_170001_909_full.fits'   /

DATATAGS= '"ESA", "NASA", "ESA/NASA"'/ Additional information
```

Note that e.g., **DETECTOR**, **GRATING**, and **FILTER** this might seem unnecessary for instruments with an *a priori* single fixed value, but for ground-based observatories, upgrades of an instrument might include a change of e.g., filters.

## 15.6. Mandatory keywords for spectrographs and filter instruments (Sections 3.2 and 5.4)

```
WAVEUNIT=                   -10 / Wavelength related kwds have unit: 10^(WAVEUNIT) m
WAVEREF = 'vac     '            / Wavelength related kwds in vacuum

WAVEMIN =                582.10 / [Angstrom] Min wavelength covered by filter
WAVEMAX =                586.62 / [Angstrom] Max wavelength covered by filter
```

For spectrographs, and narrow-band filter instruments (when radial velocity is of importance when interpreting the observations), the following keyword is mandatory:

```
OBS_VR  =                 36.62 / [km/s] Observer's outward velocity w.r.t. Sun
```

Also, to signal that no wavelength correction has been applied (even if the observer moves with respect to the Sun with a velocity given by **OBS_VR**) the following WCS keywords are mandatory:

```
SPECSYS = 'TOPOCENT'            / Coordinate reference frame = observer
VELOSYS =                   0.0 / [m s-1] No velocity correction applied to WAVE coord.
```

## 15.7. Mandatory keyword for spectrographs (Section 5.4)

```
SLIT_WID=                   0.5 / [arcsec] Slit width
```

## 15.8. Mandatory keywords for polarimetric data (Section 5.4.1)

```
POLCCONV= '(+HPLT,-HPLN,+HPRZ)' / Reference system for Stokes vectors
POLCANGL=                  45.5 / [deg] Counter-clockwise rotation around +HPRZ axis
```

## 15.9. Mandatory keyword for grouping (Sections 7 and Appendix V-b )

```
POINT_ID= '20201224_165812_200'/ Unique (re-)pointing ID
```

# 16. Mandatory keyword for SOLARNET HDUs that contain keywords with a definition in conflict with the specifications in this document (Section 2.2)

If a SOLARNET HDU contains SOLARNET keywords with definitions that are in conflict with the definitions in this document, those keywords *must* be listed as a comma-separated list in the keyword **SOLNETEX**, e.g.:

```
SOLNETEX= 'PLANNER, ATMOS_R0'  / Exception: kws with conflicting definitions
```

Note that *none* of the keywords that are mandatory for a given HDU may have conflicting definitions[17].

# 17. Mandatory keywords for all HDUs that uses any of the variable-keyword, pixel list or meta-observation mechanism (Sections 2.1, 2.2, 2.3, Appendix I, Appendix II, and Appendix III)

Any HDU using one of these mechanisms must have **SOLARNET** set to a non-zero value, even non-Obs-HDUs (which should use a value of -1). In addition, **EXTNAME** must be set according to the guidelines in Section 2. Finally, the respective **VAR_KEYS**, **PIXLISTS** or **METADIM**/**METAFILS** must be set – see Appendix I, Appendix I-d and Appendix III for details.

```
SOLARNET =                -1.0 / SOLARNET mechanisms may be used
EXTNAME = 'zunhousekeeping'    / Name of HDU
VAR_KEYS= 'He_I_T3;TEMPS'      / Variable keyword used by this Aux-HDU
PIXLISTS= 'He_I_T3;T_IDX'      / Pixel list used by this Aux-HDU
METADIM =                    4 / Split dimension of meta-observation
METAFILS= 'sleep_a_zun_l2_20201224_165812.2_balanced.fits,                &'
CONTINUE  'sleep_a_zun_l2_20201224_165912.4_balanced.fits,                &'
CONTINUE  'sleep_a_zun_l2_20201224_170012.6_balanced.fits,                &'
CONTINUE  'sleep_a_zun_l2_20201224_170112.8_balanced.fits,                &'
CONTINUE  ''  / All files in Meta-obs
```

## 17.1. Mandatory keyword for binary table extension value columns that use pixel-to-pixel association (Appendix I-b)

The value of **WCSNn** must start with "**PIXEL-TO-PIXEL**" to signal that a direct pixel-to-pixel association applies:

```
WCSN5   = 'PIXEL-TO-PIXEL'       / Column 5 uses pixel-to-pixel association
```

[17] If some existing utility requires a different definition of a mandatory keyword, we recommend that the value for the non-SOLARNET definition be given in a new keyword, and that the software be modified.

## 17.2. Mandatory keywords for binary table extensions that use the SOLARNET pixel lists mechanism for flagging pixels (Appendix I-d)

Binary table columns storing pixel indices must have `TCTYPn` equal to `'PIXEL'` and `TTYPEn` equal to `'DIMENSIONk'`, where `k` is the dimension number in the referring HDU's data cube. The column storing pixel types must have `TTYPEn` equal to `'PIXTYPE'`. Any attribute columns must have `TTYPEn` equal to the name of the attached attribute contained in that column. E.g.:

```
TTYPE1   = 'DIMENSION1'         / Col.1 is index into data cube dimension 1
TTYPE2   = 'DIMENSION2'         / Col.2 is index into data cube dimension 2
TTYPE3   = 'DIMENSION3'         / Col.3 is index into data cube dimension 3
TTYPE4   = 'PIXTYPE'            / Col.4 is the pixel type
TTYPE5   = 'ORIGINAL'           / Col.5 contains original values of listed pixels
TCTYP1   = 'PIXEL   '           / Indicates that col. 1 is a pixel index
TCTYP2   = 'PIXEL   '           / Indicates that col. 2 is a pixel index
TCTYP3   = 'PIXEL   '           / Indicates that col. 3 is a pixel index
```

# *18. Optional keywords for all Obs-HDUs*

The keywords in this section are optional for both fully and partially SOLARNET-compliant Obs-HDUs.

## 18.1. Optional keyword for deep space missions (not Earth orbiting satellites) (Section 3.2)

In addition to the mandatory keyword **`DSUN_OBS`** the optional keyword **`DSUN_AU`** may be used to give the instrument-Sun centre distance in astronomical units:

```
DSUN_AU =       0.594805136980 / [AU] Distance from instrument to Sun centre)
```

## 18.2. Optional date and time keywords (Section 4)

```
DATE-END= '2020-12-24T17:00:02.5' / Date of end of observation
DATE-AVG= '2020-12-24T17:00:01.3' / Average date of observation
TIMESYS = 'UTC     '              / Time scale of the time-related keywords.
```

## 18.3. Optional keywords describing cadence (Section 5.3)

```
CADENCE =                  2.5 / [s] Planned cadence
CADAVG  =              2.45553 / [s] Average actual cadence
CADMIN  =              2.27943 / [s] Minimum actual frame-to-frame spacing
CADMAX  =              2.69162 / [s] Maximum actual frame-to-frame spacing
CADVAR  =            0.0118546 / [s] Variance of frame-to-frame spacing
```

## 18.4. Optional keyword for spectrographs (Section 5.4)

```
WAVECOV = '70.351919-70.839504, 76.749034-77.236619, 77.743708-78.231293, 78.3&'
CONTINUE  '67817-78.991925, 97.496379-97.842807, 102.2309-102.82752, 102.99111&'
CONTINUE  '-103.31829, 103.4049-104.00153&' / [nm] All WAVEMIN-WAVEMAX
```

## 18.5. Optional data description keywords (Section 5.1)

```
BNAME    = 'Spectral Radiance'    / Description of what the data array represents
UCD      = 'phot.radiance;em.UV' / Unified Content Descriptors v1.23
```

## 18.6. Optional keywords characterizing the instrument/data (Sections, 5.4 and 5.5)

```
WAVEBAND= 'He I 584.58 A'       / Strongest emission line in data
WAVELNTH=               584.61 / [Angstrom] Characteristic wavelength in vacuum
RESOLVPW=                 6530 / Resolving power of spectrograph

FT_LOCK =                    1 / Feature tracking on
ROT_COMP=                    1 / Solar rotation compensation on (1)/off (0)
ROT_MODL= 'SNODGRASS'          / Model used for rotation compensation
ROT_FORM= 'A+B*sin^2(phi)+C*sin^4(phi)' / [deg/day] Snodgrass, doi:10.1086/168467

AO_LOCK =                  0.9 / Adaptive optics status, 0.0=no lock, 1.0=lock
AO_NMODE=                    2 / Type of modes: Zernike, Karhunen-Loeve
```

For radio observations:

```
BNDCTR =                   198 / [GHz] Characteristic frequency in vacuum
```

The response function may be given in the variable keyword **RESPONSE**. If the data has been corrected for a variable response, the response function that has been applied should instead be given in **RESPAPPL**:

```
RESPONSE=                 0.76 / Mean of response function
```

or

```
RESPAPPL=                 0.52 / Mean of applied response function
```

## 18.7. Optional quality aspects keywords (Sections 3.1 and 5.5)

```
ATMOS_R0=                   15 / [cm] Atmospheric coherence length
ELEV_ANG=                 76.2 / [deg] Telescope elevation angle.
OBS_LOG = 'logs/2020/12/24/'   / URL of observation log
RSUN_REF=          6.95508E+08 / [m] Solar rad. used for calc. px scale
COMPQUAL=                 0.75 / Quality of data after lossy compression
COMP_ALG= 'jpeg2000'           / Name of lossy compression algorithm
```

## 18.8. Optional data statistics keywords (Section 5.6)

```
DATAMIN =                  -23 / [DN] Minimum of data
DATAMAX =                 4077 / [DN] Maximum of data
DATAMEAN=              144.794 / [DN] Mean of data
DATAMEDN=              11.0000 / [DN] Median of data
DATAP01 =                  -14 / [DN] 1st  percentile of data
DATAP10 =                   -9 / [DN] 10th percentile of data
DATAP25 =                   -4 / [DN] 25th percentile of data
```

```
DATAP75 =                   70 / [DN] 75th percentile of data
DATAP90 =                  304 / [DN] 90th percentile of data
DATAP95 =                  704 / [DN] 95th percentile of data
DATAP98 =                 2085 / [DN] 98th percentile of data
DATAP99 =                 2844 / [DN] 99th percentile of data
DATANP01=            -0.099689 / DATAP01/DATAMEAN
...
DATANP99=              19.6417 / DATAP99/DATAMEAN
DATARMS =              471.524 / [DN] RMS dev: sqrt(sum((data-DATAMEAN)^2)/N)
DATANRMS=              3.25652 / DATARMS/DATAMEAN
DATAMAD =              3.43345 / [DN] Mean abs dev: sum(abs(data-DATAMEAN))/N
DATANMAD=           0.02371265 / DATAMAD/DATAMEAN
DATAKURT=              24.8832 / Kurtosis of data
DATASKEW=              4.84719 / Skewness of data
```

## 18.9. Optional keywords for missing and saturated pixels (Section 5.6.1)

```
NTOTPIX =               262144 / Expected number of data pixels
NLOSTPIX=                  512 / Number of lost pix. b/c acquisition problems
NSATPIX =                    4 / Number of saturated pixels
NSPIKPIX=                    7 / Number of noise spike pixels
NMASKPIX=                   71 / Number of masked pixels
NAPRXPIX=                   64 / Number of approximated pixels
NDATAPIX=               261550 / Number of usable pix. excl lost/miss/spik/sat
PCT_LOST=             0.195312 / NLOSTPIX/NTOTPIX*100
PCT_SATP=           0.00152588 / NSATPIX/NTOTPIX*100
PCT_SPIK=           0.00267029 / NSPIKPIX/NTOTPIX*100
PCT_MASK=           0.02708435 / NMASKPIX/NTOTPIX*100
PCT_APRX=           0.02441406 / NAPRXPIX/NTOTPIX*100
PCT_DATA=           99.7734070 / NDATAPIX/NTOTPIX*100
```

## 18.10. Optional pipeline processing keywords (Sections 8, 8.1 and 8.2)

```
LEVEL   = '3       '           / Data level of fits file
VERSION =                    2 / FITS file processing generation/version

CREATOR = 'ZUN_MOMF_PIPELINE'  / Name of software pipeline that produced the FITS file
VERS_SW = '2.5'                / Version of CREATOR software applied
HASH_SW = ' a7ef89ad998ea7feef4bbc0bbc1bbc2bbc3bbc4'/ GIT commit hash for pipeline
VERS_CAL= '2.4'                / Version of calibration pack applied

PRSTEP1 = 'MOMFBD  '           / Processing step type
PRPROC1 = 'zun_momf.pro'       / Name of procedure performing PRSTEP1
PRPVER1 =                  1.5 / Version of procedure PRPROC1
PRMODE1 = 'BALANCED'           / Processing mode of PRPROC1
PRPARA1 = 'ITER=5,MANUAL=1'    / List of parameters/options for PRPROC1
PRREF1  = 'miss.influencer@esa.int' / Factors influencing PRSTEP1
PRLOG1  = ' % Program caused arithmetic error: Integer divide by 0' / PRPROC1 log
PRENV3  = '  Kernel: Linux                                                    &'
CONTINUE  '  Kernel release number: 3.10.0-1160.36.2.el7.x86_64               &'
CONTINUE  '  OS: Red Hat Enterprise Linux Server release 7.9 (Maipo)          &'
CONTINUE  '  CPU: Intel(R) Xeon(R) CPU E5-2630L v4 @ 1.80GHz                  &'
CONTINUE  '  IDL 8.5 (Jul  7 2015), memory bits: 64, file offset bits: 64     &'
CONTINUE  ''  / Processing environment of PRSTEP1

PRLIB1A = 'ZUNRED     '        / Software library containing PRPROC1
PRVER1A =                32214 / Version of PRLIB1A
PRHSH1A = 'a7ef89ad998ea7feef4bbc0bbc1bbc2bbc3bbc4' / GIT commit hash for PRLIB1A
PRBRA1A = 'production'         / GIT/SVN repository branch of PRLIB1A
```

```
PRLIB1B = 'SSW/vobs/ontology/idl/gen_temp,SSW/packages/sunspice/idl/atest,SSW/&'
CONTINUE  'so/spice/idl/atest,SSW/vobs/gen/idl,SSW/soho/gen/idl/util,SSW/gen/i&'
CONTINUE  'dl_libs/astron/coyote' / Software library containing PRPROC1
PRVER1B =                59549 / Modified Julian date of last mirroring of PRLIB1B
```

## 18.11. Optional keyword for administrative information (Section 9)

```
INFO_URL= 'http://sleep.esa.int/zun/info.html' / Data set resource web page
RELEASE = '2022-08-25T00:00'   / Public release date of data
RELEASEC= 'embargo@zun.no,zunteam@esa.int' / Data release administrators
```

## 18.12. Optional keywords for grouping (Section 7)

```
SVO_SEP1= 'POINT_ID,INSTRUME,DETECTOR,FILTER,NBIN' / Most fine grained separation
SVO_SEP2= ‘POINT_ID,INSTRUME,DETECTOR,FILTER’ / Img. shows target even w/binning
SVO_SEP3= ‘POINT_ID,INSTRUME,DETECTOR’ / Target identifiable in all filters
SVO_SEP4= ‘POINT_ID,INSTRUME’ / Still useful

SVO_GRP = 'R_SMALL_HRES_MCAD_Polar-Observations' / SVO group
MOSAICID= '10023b_2' / Mosaic ID
```

## 18.13. Optional keyword for binary table extensions using the variable-keyword, pixel list or meta-observation mechanism (Appendix I, Appendix I-d and Appendix III)

`TDESCn` may be used to give a description of the contents of the binary table column.

```
TDESC5  = 'Atmospheric coherence length values stored in this column are obtain&'
CONTINUE  'ned using the R0SUPER instrument, SW version 1.3'
```

# *Part C. Alphabetical listings of FITS keywords with section references*

# 19. Alphabetical listing of all new SOLARNET keywords with section references

Below is an alphabetical listing of all SOLARNET keywords that are not part of the FITS standard or any widely accepted FITS convention, keywords that have been used in the past that do not have widely accepted definitions, or previously defined keywords that need to take special values in SOLARNET files:

```
ANA_NCMP      Appendix IX.
AO_LOCK       5.5., 18.6.
AO_NMODE      5.5., 18.6.
ATMOS_R0      5.5., Appendix I., Appendix I-a., Appendix I-b., 16., 18.7.
BNAME         5.1., 15.4., 18.5.
BNDCTR        5.4., 18.6.
BTYPE         5.1., 15.4.
CADAVG        5.3., 18.3.
CADENCE       5.3., 18.3.
CADMAX        5.3., 18.3.
CADMIN        5.3., 18.3.
CADVAR        5.3., 18.3.
CCURRENT      Appendix V-b., 15.5.
CMPDESn       Appendix IX.
CMPINCn       Appendix IX.
CMPMULn       Appendix IX.
CMPNAMn       Appendix IX.
CMPSTRn       Appendix IX.
CMPTYPn       Appendix IX.
CMP_NPn       Appendix IX.
COMPQUAL      5.5., 18.7.
COMP_ALG      5.5., 18.7.
CONSTEXT      Appendix IX.
CPDISia       Appendix VI.
CPERRia       Appendix VI.
CWDISia       Appendix VI.
CWERRia       Appendix VI.
DATAEXT       Appendix IX.
DATAMAD       5.6., 18.8.
DATANMAD      5.6., 18.8.
DATANPnn      5.6., 18.8.
DATANRMS      5.6., 18.8.
DATAPnn       5.6., 18.8.
DATATAGS      6., 15.5.
DSUN_AU       3.2., 18.1.
DWia          Appendix VI., Appendix VI-a.
ELEV_ANG      5.5., 18.7.
FILEVERP      Appendix V-a.
FT_LOCK       5.5., 18.6.
HASH_SW       8.1., 18.10.
INCLEXT       Appendix IX.
INFO_URL      9., 18.11.
METADIM       Appendix III., Appendix III-a., 17.
METADIMn      Appendix III-a.
METAFILS      Appendix III., 17.
MISSION       6., 15.5.
MOSAICID      7., 18.12.
NAPRXPIX      5.6.1., 18.9.
NBIN          5.2., 15.4.
NBINj         5.2., 15.4.
NDATAPIX      5.6.1., 18.9.
NLOSTPIX      5.6.1., 18.9.
```

| | |
|---|---|
| **TELCONFG** | **6., 15.5.** |
| **TEXPOSUR** | **5.2., 15.4.** |
| **TKEYSn** | **Appendix IV.** |
| **TPXLSn** | **Appendix IV.** |
| **TVARKn** | **Appendix IV.** |
| **VAR_KEYS** | **Appendix I., Appendix I-d., Appendix IV., Appendix VII., 17.** |
| **VERSION** | **8., 18.10.** |
| **VERS_CAL** | **8.1., 18.10.** |
| **VERS_SW** | **8.1., 18.10.** |
| **WAVEBAND** | **5.4., 18.6.** |
| **WAVECOV** | **5.4., 18.4.** |
| **WAVEMAX** | **5.4., 15.6., 18.4.** |
| **WAVEMIN** | **5.4., 15.6.** |
| **WAVEREF** | **5.4., 15.6.** |
| **WGTEXT** | **Appendix IX.** |
| **XDIMTYm** | **Appendix IX.** |
| **XDIMXTm** | **Appendix IX.** |

**Number of keywords: 130**

# *20. Alphabetical listing of all keywords with section references*

Below is an alphabetical listing of all keywords used in this document, both the SOLARNET keywords listed in Section 19 and FITS standard/widely accepted FITS convention keywords, the latter identified with a one-character trailing code:

- **S**: Keywords defined in the FITS standard (listed in the FITS standard Appendix C)
- **P**: Keywords defined in Paper I-V and Thompson (2006)[18].
- **O**: Other keywords in common use before being specified in this document

| | | |
|---|---|---|
| **ANA_NCMP** | | **Appendix IX.** |
| **AO_LOCK** | | **5.5., 18.6.** |
| **AO_NMODE** | | **5.5., 18.6.** |
| **ATMOS_R0** | | **5.5., Appendix I., Appendix I-a., Appendix I-b., 16., 18.7.** |
| **AUTHOR** | **S** | **6., 15.5.** |
| **BLANK** | **S** | **5.6.2., 15.4.** |
| **BNAME** | | **5.1., 15.4., 18.5.** |
| **BNDCTR** | | **5.4., 18.6.** |
| **BTYPE** | | **5.1., 15.4.** |
| **BUNIT** | **S** | **5.1., 15.4.** |
| **BZERO** | **S** | **Appendix IV.** |
| **CADAVG** | | **5.3., 18.3.** |
| **CADENCE** | | **5.3., 18.3.** |
| **CADMAX** | | **5.3., 18.3.** |
| **CADMIN** | | **5.3., 18.3.** |

[18] Some of the WCS keywords defined in these papers are also defined in the FITS standard Table 22

**Number of keywords: 216**